%% file: main.tex
\documentclass[sigplan, 10pt]{acmart}
\usepackage{booktabs} 

\input{header}

\begin{document}
	\title{\sys{}: A Data Store for Analytics on Large Videos}
	
	%
	%
	
	\author{Tiantu Xu}
	\affiliation{%
		\institution{Purdue ECE}
	}
	\author{Luis Materon Botelho}
	\affiliation{%
		\institution{Purdue ECE}
	}
	
	\author{Felix Xiaozhu Lin}
	\affiliation{%
		\institution{Purdue ECE}
	}
	
	\input{abstract}
	
	\copyrightyear{2019} 
	\acmYear{2019} 
	\setcopyright{acmcopyright}
	\acmConference[EuroSys '19]{Fourteenth EuroSys Conference 2019}{March 25--28, 2019}{Dresden, Germany}
	\acmBooktitle{Fourteenth EuroSys Conference 2019 (EuroSys '19), March 25--28, 2019, Dresden, Germany}
	\acmPrice{15.00}
	\acmDOI{10.1145/3302424.3303971}
	\acmISBN{978-1-4503-6281-8/19/03}

	
		
	\settopmatter{printfolios=true}
	\maketitle
	
	\input{intro}   	
	\input{bkgnd}

\input{case}

	\input{istream}

	\input{impl}		
	\input{eval-sys}
\input{discussion}

	\input{related}
\input{conclusion}      
	\input{ack}

	
	\bibliographystyle{ACM-Reference-Format}
	\input{vstore.bbl}

\end{document}

%% file: header.tex
\usepackage{wrapfig}
\usepackage{comment}
\usepackage{colortbl}

\usepackage{hyperref}
\usepackage{tabularx}

\usepackage[T1]{fontenc} 

\usepackage{color,soul}
\usepackage{graphics}
\usepackage{lipsum}

\usepackage{tikz}
\usetikzlibrary{decorations.pathmorphing, patterns,shapes,snakes} 

\usepackage[inline]{enumitem}	
\usepackage{listings} 

\usepackage{capt-of}	

\ifdefined\HCode
\usepackage[compatibility=false]{caption}

\else
\usepackage{subcaption} 
\fi

\usepackage{mathrsfs}	
\usepackage{amsfonts}

\usepackage[export]{adjustbox} 

\usepackage{booktabs}

\definecolor{myForestGreen}{RGB}{34, 139, 34}
\makeatletter
\def\blfootnote{\xdef\@thefnmark{}\@footnotetext}
\makeatother

\definecolor{airforceblue}{rgb}{0.36, 0.54, 0.66}
\newcommand*\circled[1]{\tikz[baseline=(char.base)]{
		\node[shape=circle,fill=black,draw,text=white,inner sep=1pt] (char) {#1};}}

\definecolor{refkey}{RGB}{34,139,34}
\definecolor{labelkey}{rgb}{0,0,1}

\newcommand{\Paragraph}[1]{\vskip 3pt\noindent\textbf{#1 }}	 

  {\begin{list}{$\bullet$}%
     {\setlength{\parsep}{0pt}%
      \setlength{\topsep}{0pt}%
      \setlength{\itemsep}{2pt}}}%
  {\end{list}}
\newcommand\note[1]{} 

\newcommand\noted[1]{} 

\newcommand\sect[1]{Section~\ref{sec:#1}}	
		
\newenvironment{myitemize}%
  {\begin{itemize}
	[leftmargin=0cm,
		itemindent=.3cm,
		labelwidth=\itemindent,
		labelsep=0pt,
		parsep=3pt,
		topsep=2pt,
		itemsep=1pt,
		align=left]
  }%
  {\end{itemize}}    

  {\begin{enumerate}
	[leftmargin=0cm,itemindent=.5cm,labelwidth=\itemindent,
		labelsep=0pt,
		parsep=1pt,
		topsep=1pt,
		itemsep=3pt,
		align=left]
  }%
  {\end{enumerate}}    
  
\newenvironment{enumerateinline}%
  {\begin{enumerate*}
	[label=\roman*)]
  }%
  {\end{enumerate*}}      

\input{terms}

\makeatletter
\def\@copyrightspace{\relax}
\makeatother

%% file: terms.tex
\newcommand{\sys}{VStore}

\newcommand{\fs}{storage format}
\newcommand{\fc}{consumption format}

\newcommand{\FS}{SF}
\newcommand{\FC}{CF}

\newcommand{\sink}{consumer}

\newcommand{\sfg}{SF\textsubscript{g}}

%% file: abstract.tex

\begin{abstract}
	
We present \sys{},
a data store for supporting fast, resource-efficient analytics over large archival videos.
\sys{} manages video ingestion, storage, retrieval, and consumption.
It controls video formats along the video data path. 
It is challenged by i) the huge combinatorial space of video format knobs; ii) the complex impacts of these knobs and their high profiling cost; iii) optimizing for multiple resource types. 
It explores an idea called backward derivation of configuration: in the opposite direction along the video data path, \sys{} passes the video quantity and quality expected by analytics backward to retrieval, to storage, and to ingestion.
In this process, \sys{} derives an optimal set of video formats, optimizing for different resources in a progressive manner.

\sys{} automatically derives large, complex configurations consisting of more than one hundred knobs over tens of video formats. 
In response to queries, \sys{} selects video formats catering to the executed operators and the target accuracy. 
It streams video data from disks through decoder to operators. 
It runs queries as fast as 362$\times$ of video realtime.

\end{abstract}

%% file: intro.tex

%
%

\newcommand{\isolation}{Enforcing isolation with ARM TrustZone}
\newcommand{\noconcurrency}{Exposing low-level abstractions of trusted computations}
\newcommand{\seqmm}{Unbounded arrays as the universal memory abstraction}
\newcommand{\hints}{Exploiting performance hints from the untrusted}

\section{Introduction}
\label{sec:intro}

Pervasive cameras produce videos at an unprecedented rate.
Over the past 10 years, the annual shipments of surveillance cameras grow by 10$\times$, to 130M per year~\cite{ihs-report-2018}. 
Many campuses are reported to run more than 200 cameras  24$\times$7~\cite{seagatereport}. 
In such deployment, a single camera produces as much as 24 GB encoded video footage per day (720p at 30 fps).

%

\Paragraph{Retrospective video analytics}
To generate insights from enormous video data, retrospective analytics is vital: video streams are captured and stored on disks for a user-defined lifespan; users run queries over the stored videos on demand. 
Retrospective analytics offers several key advantages that live analytics lacks.
i) Analyzing many video streams in real time is expensive, e.g., 
running deep neural networks over live videos from a \$200 camera may require a \$4000 GPU~\cite{noscope}.
ii) Query types may only become known after the video capture~\cite{blazeit}. 
iii) At query time, users may interactively revise their query types or parameters~\cite{blazeit,eva},
which may not be foreseen at ingestion time. 
iv) In many applications, only a small fraction of the video will be eventually queried~\cite{ihs-report-2016}, making live analytics an overkill.


A video query (e.g., ``what are the license plate numbers of all blue cars in the last week?'') is typically executed as a cascade of operators~\cite{shen17cvpr,chameleon,blazeit,noscope,videostorm}.
Given a query, a query engine assembles a cascade and run the operators. 
Query engines typically expose to users the trade-offs between operator accuracy and resource costs, allowing users to obtain inaccurate results with a shorter wait. 
The users thus can explore large videos interactively~\cite{blazeit,noscope}. 
Recent query engines show promise of high speed, e.g., consuming one-day video in several minutes~\cite{noscope}.


\input{fig-concept}

\Paragraph{Need for a video store}
While recent query engines assume \textit{all} input data as raw frames present in memory,
there lacks a video store that manages large videos for analytics. 
The store should orchestrate four major stages on the video data path: 
ingestion, storage, retrieval, and consumption,
as shown in Figure~\ref{fig:concept}. 
The four stages demand multiple hardware resources, including 
encoder/decoder bandwidth, 
disk space, and CPU/GPU cycles for query execution. 
The resource demands are high, thanks to large video data.
Demands for different resource types may conflict. 
Towards optimizing these stages for resource efficiency, 
classic video databases are inadequate~\cite{kang09ms}: 
they were designed for \textit{human} consumers watching videos at 1$\times$--2$\times$ speed of video realtime; 
they are incapable of serving some \textit{algorithmic} consumers, i.e., operators, processing videos at more than 1000$\times$ video realtime. 
Shifting part of the query to ingestion~\cite{focus} has important limitations and does not obviate the need for such a video store, as we will show in the paper. 


Towards designing a video store, 
we advocate for taking a key opportunity: 
as video flows through its data path, the store should control video formats (fidelity and coding) through extensive video parameters called \textit{knobs}. 
These knobs have significant impacts on resource costs and analytics accuracy, opening a rich space of trade-offs.

We present \sys{}, a system managing large videos for retrospective analytics. 
The primary feature of \sys{} is its automatic configuration of video formats.  
As video streams arrive, \sys{} saves multiple video versions and judiciously sets their \textit{\fs{}s};
in response to queries, \sys{} retrieves stored video versions and converts them into \textit{\fc{}s} catering to the executed operators.
Through configuring video formats,  
\sys{} ensures operators to meet their desired accuracies at high speed; 
it prevents video retrieval from bottlenecking consumption; 
it ensures resource consumption to respect budgets. 

To decide video formats,  
\sys{} is challenged by 
i) an enormous combinatorial space of video knobs;
ii) complex impacts of these knobs and high profiling costs; 
iii) optimizing for multiple resource types. 
These challenges were unaddressed:
while classic video databases may save video contents in multiple formats, 
their format choices are oblivious to analytics and often ad hoc~\cite{kang09ms};
while existing query engines recognize the significance of video formats~\cite{blazeit,videostorm,chameleon} and optimize them for query execution, 
they omit video coding, storage, and retrieval, which are all crucial to retrospective analytics.

To address these challenges, 
our key idea behind \sys{} is \textit{backward derivation}, shown in Figure~\ref{fig:concept}. 
In the opposite direction of the video data path,
\sys{} passes the desired data quantity and quality from algorithmic consumers backward to retrieval, to storage, and to ingestion. 
In this process, \sys{} optimizes for different resources in a progressive manner;
it elastically trades off among them to respect resource budgets. 
More specifically, 
i) from operators and their desired accuracies, \sys{} derives video formats for fastest data consumption, for which it effectively searches in a high-dimensional parameter space with video-specific heuristics;
ii) from the \fc{}s, \sys{} derives video formats for storage, for which 
it systematically coalesces video formats to optimize for ingestion and storage costs; 
iii) from the \fs{}s, \sys{} derives a data erosion plan, which gradually deletes aging video data, trading off analytics speed for lower storage cost.

Through evaluation with two real-world queries over six video datasets, 
we demonstrate that \sys{} is capable of deriving large, complex configuration with hundreds of knobs over tens of video formats, which are infeasible for humans to tune. 
Following the configuration, \sys{} stores multiple formats for each video footage. 
To serve queries, it streams video data (encoded or raw) from disks through decoder to operators, 
running queries as fast as 362$\times$ of video realtime. 
As users lower the target query accuracy, \sys{} elastically scales down the costs by switching operators to cheaper video formats, accelerating the query by two orders of magnitude. 
This query speed is 150$\times$ higher compared to systems that lack automatic configuration of video formats. 
\sys{} reduces the total configuration overhead by 5$\times$.


\Paragraph{Contributions}
We have made the following contributions. 

\begin{myitemize}
\item We make a case for a new video store for serving retrospective analytics over large videos. 
We formulate the design problem and experimentally explore the design space. 

\item 
To design such a video store, 
we identify the configuration of video formats as the central concern. 
We present a novel approach called backward derivation. 
With this approach, we contribute new techniques for searching large spaces of video knobs, for coalescing stored video formats, and for eroding aging video data. 

\item We report \sys{}, a concrete implementation of our design. 
Our evaluation shows promising results. 
\sys{} is the first holistic system that manages the full video lifecycle optimized for retrospective analytics, to our knowledge. 
\end{myitemize}

%% file: fig-concept.tex

\ifpdf

\begin{figure}[t!]
\centering

   \includegraphics[width=.9\linewidth]{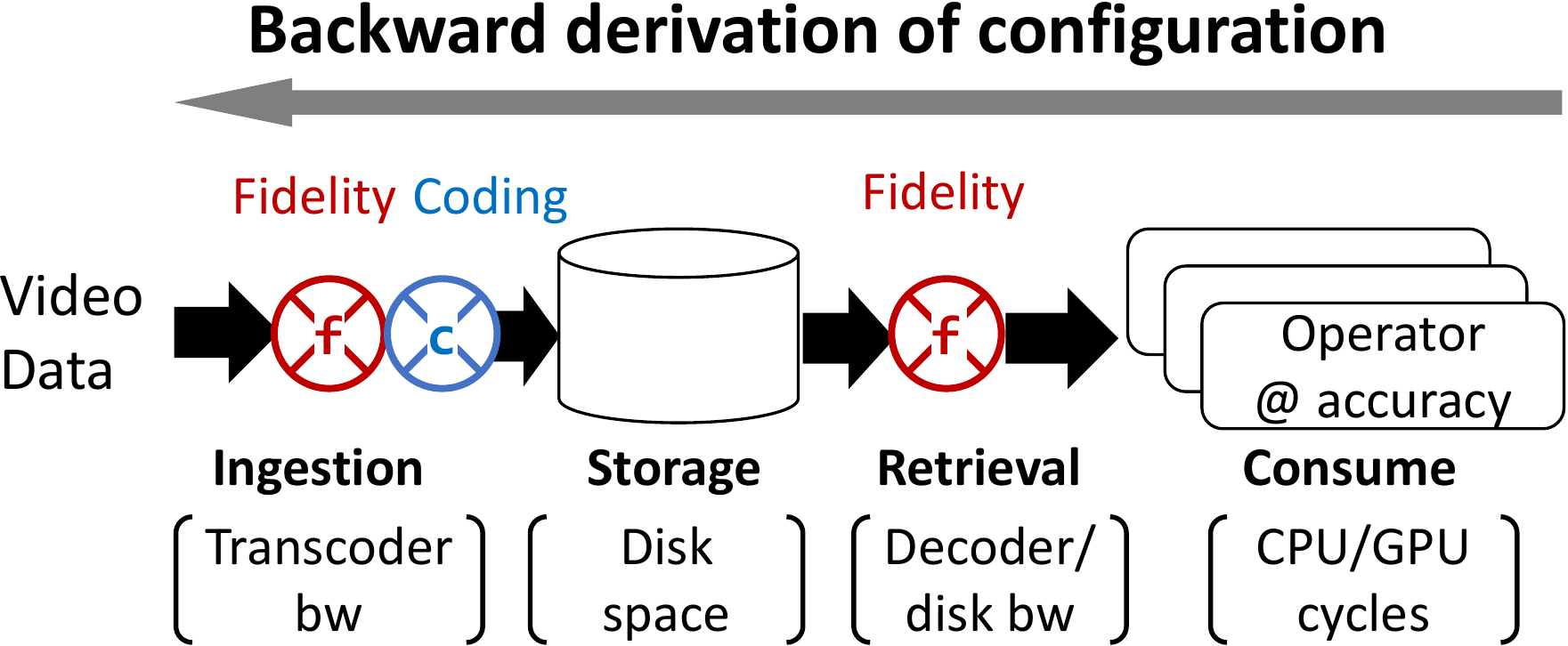}
	

\caption{The \sys{} architecture, showing the video data path and backward derivation of configuration.
}


\label{fig:concept}
\end{figure}

\fi

%% file: bkgnd.tex

\section{Motivations}
\label{sec:bkgnd}



%
%

\subsection{Retrospective Video analytics}
\label{sec:bkgnd:analytics}

\input{fig-pipeline}

\Paragraph{Query \& operators}
A video query is typically executed as a cascade of operators.
As shown in Figure~\ref{fig:pipeline},
early operators scan most of the queried video timespan at low cost.
They activate late operators over a small fraction of video for deeper analysis.
Operators consume raw video frames. 
Of a cascade, the execution costs of operators can differ by three orders of magnitude~\cite{noscope}; 
they also prefer different input video formats, catering to their internal algorithms. 

\Paragraph{Accuracy/cost trade-offs in operators}
An operator's output quality is characterized by \textit{accuracy}, i.e., how close the output is to the ground truth. 
We use a popular accuracy metric called F1 score: the harmonic mean of precision and recall~\cite{chameleon}.
At runtime, an operator's target accuracy is set in queries~\cite{chameleon, videostorm, blazeit}.
\sys{}  seeks to provision minimum resources for operators to achieve the target accuracy.
\subsection{System model}
\label{sec:bkgnd:model}

We consider a video store running on one or a few commodity servers. 
Incoming video data flows through the following major system components. 
We assume a pre-defined library of operators, the number of which can be substantial;
each operator may run at a pre-defined set of accuracy levels.
By combining the existing operators at different accuracy levels, 
a variety of queries can be assembled. 
We will discuss how operator addition/deletion may be handled in Section~\ref{sec:discussion}.

\begin{myitemize}
\item 
\textbf{Ingestion:}
Video streams continuously arrive. 
In this work, we consider the input rate of incoming video as given.
The ingestion optionally converts the video formats, e.g., by resizing frames. 
It saves the ingested videos either as encoded videos (through transcoding) or as raw frames. 
The ingestion throughput is bound by transcoding bandwidth, typically one order of magnitude lower than disk bandwidth. This paper will present more experimental results on ingestion. 
%
%
%
\item 
\textbf{Storage:}
Like other time-series data stores~\cite{summarystore}, 
videos have age-based values. 
A store typically holds video footage for a user-defined lifespan~\cite{oraclereport}.
In queries, users often show higher interest in more recent videos. 

\item 
\textbf{Retrieval:} 
In response to operator execution, the store retrieves video data from disks, optionally converts the data format for the operators, and supplies the resultant frames. 
If the on-disk videos are encoded, the store must decode them before supplying.
Data retrieval may be bound by decoding or disk read speed. 
Since the decoding throughput (often tens of MB/sec) is far below disk throughput (at least hundreds of MB/sec), the disk only becomes the bottleneck in loading \textit{raw} frames. 

\item 
\textbf{Consumption:} 
The store supplies video data to \sink{}s, i.e., operators spending GPU/CPU cycles to consume data. 

\end{myitemize}

Figure~\ref{fig:concept} summarizes the resource cost of the
components above.
The retrieval/consumption \textit{costs} are reciprocal to data retrieval/consumption \textit{speed}, respectively.
The operator runs at the speed of retrieval or consumption, whichever is lower. 
To quantify operator speed, we adopt as the metric the ratio between video duration and video processing delay.
For instance, if a 1-second video is processed in 1 ms, the speed is 1000$\times$ realtime. 
 


\Paragraph{Key opportunity: controlling video formats}
As video data flows through, a video store is at liberty to control the video formats. 
This is shown in Figure~\ref{fig:concept}. 
At the ingestion, the system decides \textit{fidelity} and \textit{coding} for each stored video version; 
at the data retrieval, the system decides the \textit{fidelity} for each raw frame sequence supplied to \sink{}s. 



\input{bkgnd-index}

\input{bkgnd-fidelity}

%% file: fig-pipeline.tex

\begin{figure}[t!]
	\centering	
	\includegraphics[width=1\linewidth]{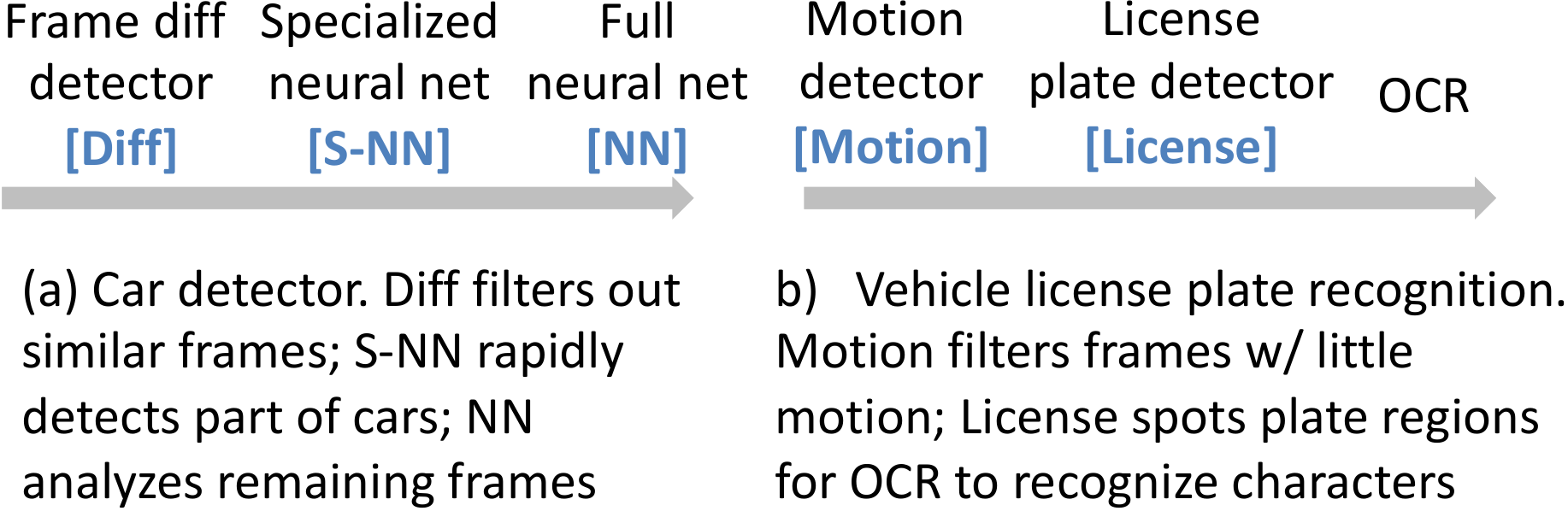}

\caption{Video queries as operator cascades~\cite{openalpr,noscope}.}
\label{fig:pipeline}
\end{figure}

%% file: bkgnd-index.tex

\Paragraph{Running operators at ingestion is not a panacea}
Recent work runs early-stage operators at ingestion to save executions of expensive operators at query time~\cite{focus}.
This approach has important limitations.
\begin{myitemize}
\item
It bakes query types in the ingestion.
Video queries and operators are increasingly rich~\cite{assari16eccv,chen18tpami,wu13cvpr,fastrcnn,fasterrcnn};
one operator (e.g., neural networks) may be instantiated with different parameters depending on training data~\cite{eureka}.
Running all possible early operators at ingestion is therefore expensive. 
\item
It bakes specific accuracy/cost trade-offs in the ingestion. 
Yet, users at query time often know better trade-offs, based on domain knowledge and interactive exploration~\cite{eva,blazeit}.
\item
It prepays computation cost for all ingested videos.
In many scenarios such as surveillance, only a small fraction of ingested video is eventually queried~\cite{eureka, seagatereport}.
As a result, most operator execution at ingestion is in vain. 
\end{myitemize}

In comparison, by preparing data for queries, a video store supports richer query types, incurs lower ingestion cost, and allows flexible query-time trade-offs.
Section~\ref{sec:discussion} will provide further discussion.

%% file: bkgnd-fidelity.tex

\subsection{Video Format Knobs}

\label{sec:bkgnd:fidelity}
\label{sec:bkgnd:knobs}


\input{tab-fidelity}

The video format is controlled by a set of parameters, or knobs. 
Table~\ref{tab:fidelity} summarizes the knobs considered in this work, chosen due to their high resource impacts. 


\Paragraph{Fidelity knobs} 
For video data, encoded or raw, 
fidelity knobs dictate i) 
the \textit{quantity} of visual information,
e.g., frame sampling which decides the frame rate; 
ii) the \textit{quality} of visual information, 
which is subject to the loss due to video compression. 
Each fidelity knob has a finite set of possible values. 
A combination of knob values constitutes a \textit{fidelity option} $\vec{f}$.
All possible fidelity options constitute a fidelity space $\mathbb{F}$. 


\Paragraph{``Richer-than'' order}
Among all possible values of one fidelity knob, 
one may establish a \textit{richer-than} order (e.g., 720p is richer than 180p).
Among fidelity \textit{options}, 
one may establish a partial order of \textit{richer-than}: 
option X is \textit{richer than} option Y 
if and only if
X has the same or richer values on all knobs and richer values on at least one knob. 
The \textit{richer-than} order does \textit{not} exist in all pairs of fidelity options, e.g., between good-50\%-720p-1/2 and bad-100\%-540p-1.
One can degrade fidelity X to get fidelity Y only if X is richer than Y.



\input{fig-knob-coding}

\Paragraph{Coding Knobs}
\label{sec:bkgnd:coding} 
Coding reduces raw video size by up to two orders of magnitude~\cite{wu18cvpr}.
Coding knobs control encoding/decoding speed and the encoded video size. 
Orthogonal to video fidelity, coding knobs provide valuable trade-offs among 
the costs of ingestion, storage, and retrieval. 
These trade-offs do not affect \sink{} behaviors -- an operator's accuracy and consumption cost. 

While a modern encoder may expose tens of coding knobs (e.g., around 50 for x264), we pick three for their high impacts and ease of interpretation.
Table~\ref{tab:fidelity} summarizes these knobs and Figure~\ref{fig:knobs-coding} shows their impacts. 
\textbf{\textit{Speed step}}
accelerates encoding/decoding at the expense of increased video size.
As shown in Figure~\ref{fig:knobs-coding}(a), it can lead up to 40$\times$ difference in encoding speed and up to 2.5$\times$ difference in storage space. 
\textbf{\textit{Keyframe interval}}:
An encoded video stream is a sequence of chunks (also called ``group of pictures''~\cite{excamera}): 
beginning with a key frame, a chunk is the smallest data unit that can be decoded independently. 
The keyframe interval offers the opportunity to accelerate decoding if the \sink{}s only sample to consume a fraction of frames.
If the frame sampling interval $N$ is larger than the keyframe interval $M$, 
the decoder can skip $N/M$ chunks between two adjacent sampled frames without decoding these chunks. 
In the example in Figure~\ref{fig:knobs-coding}(b), 
smaller keyframe intervals increase decoding speed by up to 6$\times$ at the expense of larger encoded videos. 
\textbf{\textit{Coding bypass}}:
The ingestion may save incoming videos as raw frames on disks. 
The resultant extremely low retrieval cost is desirable to some fast \sink{}s (see \sect{case}). 

A combination of coding knob values is a coding option $\vec{c}$. 
All possible coding options constitute a coding space $\mathbb{C}$.

\subsection{Knob impacts}
\label{sec:bkgnd:knob-impacts}

\input{fig-motiv}

%
As illustrated in Figure~\ref{fig:concept}:
for on-disk videos, fidelity and coding knobs jointly impact
the costs of ingestion, storage, and retrieval; 
for in-memory videos to be consumed by operators, fidelity knobs impact the consumption cost and the consuming operator's accuracy. 
We have a few observations.
\textbf{Fidelity knobs enable rich cost/accuracy trade-offs.}
As shown in Figure~\ref{fig:motiv}, one may reduce resource costs by up to 50\% 
with minor (5\%) accuracy loss.
%
%
%
\textbf{The knobs enable rich trade-offs among resource types.}
This is exemplified in Figure~\ref{fig:radar}:
although three video fidelity options all lead to similar operator accuracy (0.8), 
there is no single most resource-efficient one, e.g.,
fidelity B incurs the lowest \textit{consumption} cost, but the high \textit{storage} cost due to its high image quality. 
\textbf{Each knob has significant impacts.}
Take Figure~\ref{fig:motiv}(b) as an example: one step change to image quality reduces accuracy from 0.95 to 0.85, the storage cost by 5$\times$, and the ingestion cost by 40\%. 
\textbf{Omitting knobs misses valuable trade-offs.} 
For instance, 
to achieve high accuracy of 0.9, 
the license detector would incur 60\% more consumption cost 
when the image quality of its input video changes from ``good'' to ``bad''.
This is because the operator must consume higher \textit{quantity} of data to compensate for the lower \textit{quality}.
Yet, storing all videos with ``good'' quality requires 5$\times$ storage space.
Unfortunately, most prior video analytics systems fix the image quality knob at the default value.

\input{fig-radar}

\Paragraph{The quantitative impacts are complex.}
i) The knob/cost relations 
are difficult to capture in analytical models~\cite{videostorm}.
ii) The quantitative relations vary across operators and across video contents~\cite{chameleon}.
This is exemplified by Figure~\ref{fig:motiv} (c) and (d) that show the same knob's different impacts on two operators. 
iii) One knob's impact depends on the values of other knobs. 
Take the license detector as an example: 
as image quality worsens, the operator's accuracy becomes more sensitive to resolution changes.
With ``good'' image quality, lowering image resolution from 720p to 540p slightly reduces the accuracy, from 0.83 to 0.81; 
with ``bad'' image quality, the same resolution reduction significantly reduces the accuracy, from 0.76 to 0.52.
While prior work assumes that certain knobs have independent impacts on accuracy~\cite{chameleon}, 
our observation shows that dependency exists among a larger set of knobs. 

\Paragraph{Summary \& discussion}
Controlling video formats is central to a video store design.
The store should actively manage fidelity and coding throughout the video data path.
To characterize knob impacts, the store needs regular profiling. 

Some video analytics systems recognize the significance of video formats~\cite{blazeit,videostorm,chameleon}. 
However, they focus on optimizing query execution yet omitting other resources, such as storage, which is critical to retrospective analytics. 
They are mostly limited to only two fidelity knobs (resolution and sampling rate) while omitting others, especially coding.
As we will show, a synergy between fidelity and coding knobs is vital.

%% file: tab-fidelity.tex


\begin{table}[]



\includegraphics[width=.98\linewidth]{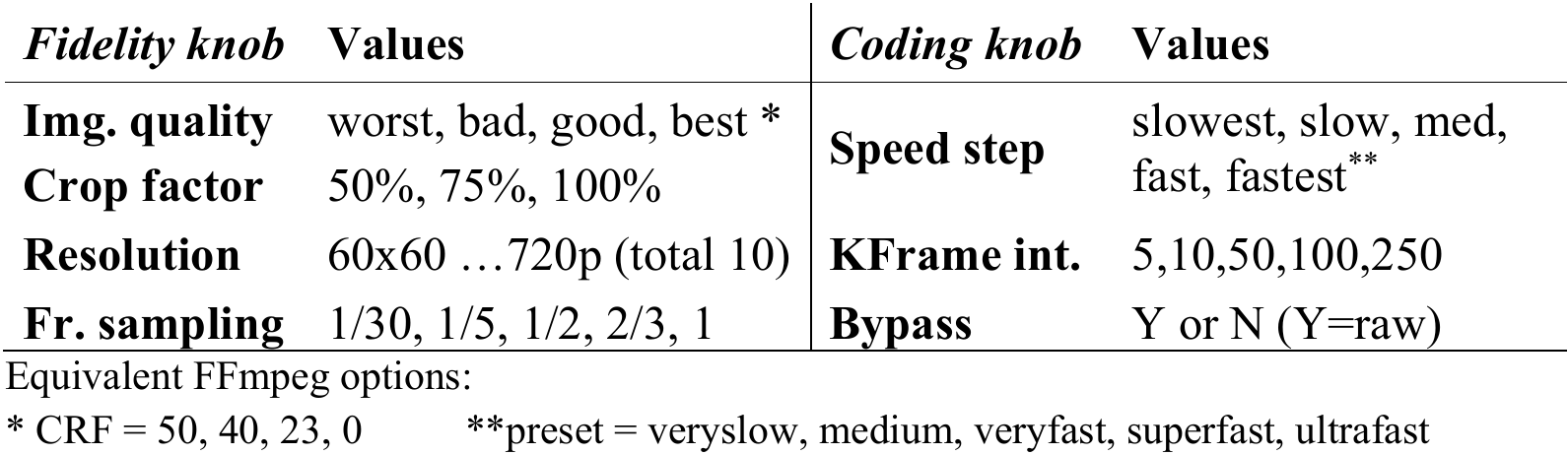}		
\centering

\caption{Knobs and their values considered in this work. 
Total: 7 knobs and 15K possible combinations of values. Note: no video quality and coding knobs for RAW.}
\label{tab:fidelity}
\end{table}

%% file: fig-knob-coding.tex

\begin{figure}[t!]
	\includegraphics[width=1\linewidth]{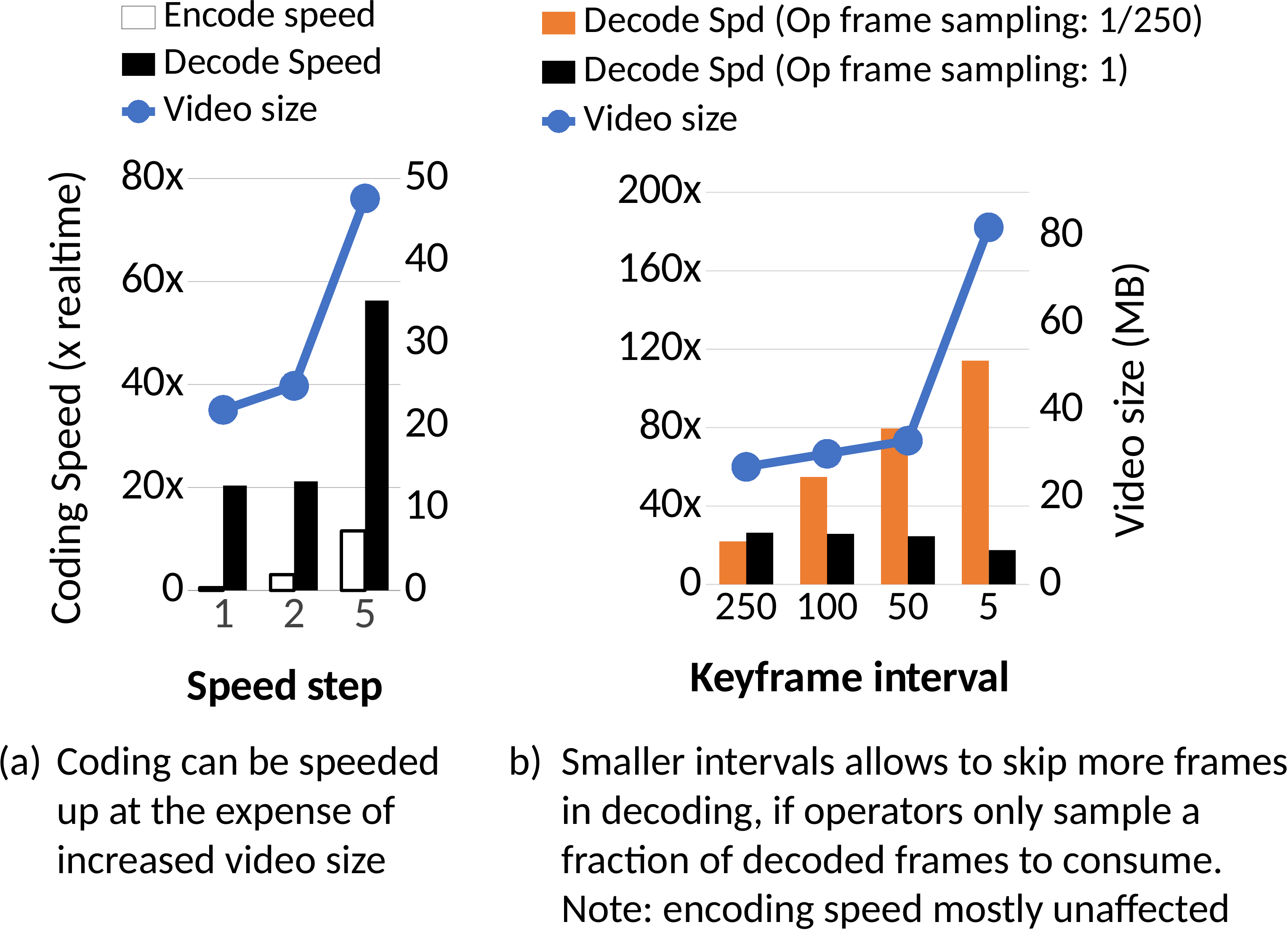}		
	\caption{Impacts of coding knobs. Video: 100 seconds from \textit{tucson}. See Section ~\ref{sec:eval} for dataset and test hardware.}
	\label{fig:knobs-coding}	
\end{figure}

%% file: fig-motiv.tex

\begin{figure}[!t]

	\includegraphics[width=1\linewidth]{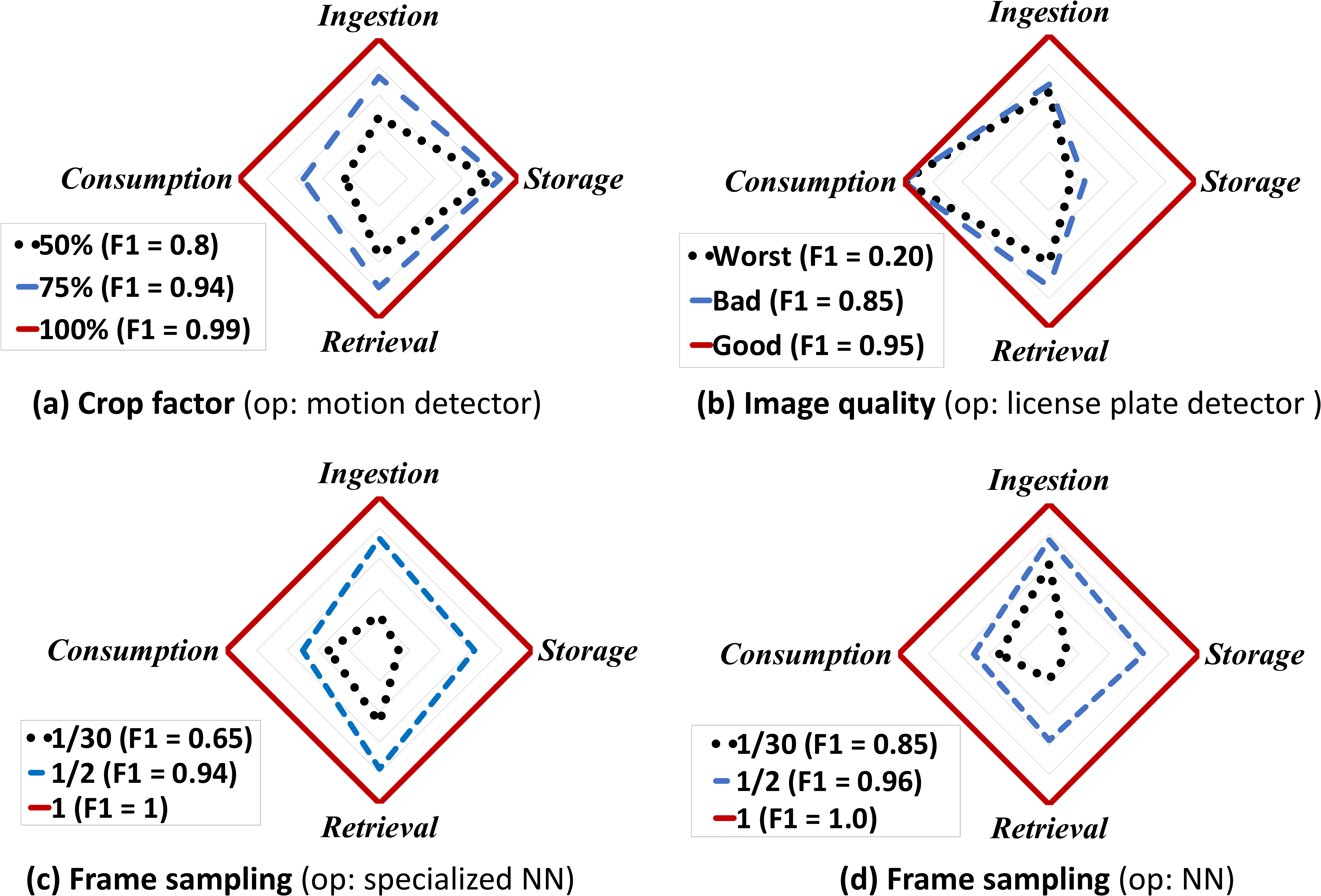}
	\caption{
	Fidelity knobs have high, complex impacts on costs of multiple components (normalized on each axis) and operator accuracy (annotated in legends). 
	Each plot: one knob changing; all others fixed. See Section ~\ref{sec:eval} for methodology.
	}
	\label{fig:motiv}
\end{figure}

%% file: fig-radar.tex

\begin{figure}[!t]

		
	\includegraphics[width=.8\linewidth]{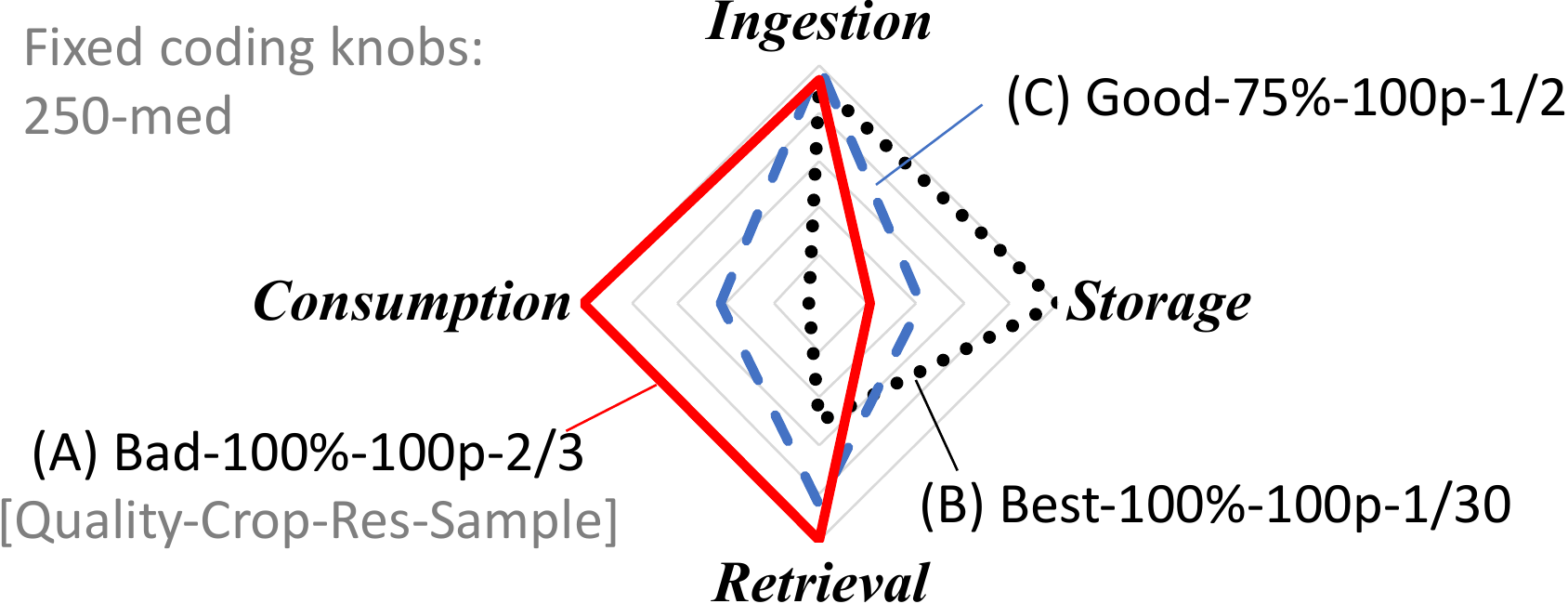}
			
	\caption{Disparate costs of fidelity options A--C, despite all leading to operator accuracy $\approx$ 0.8. Operator: License.
	Cost normalized on each axis. See Section ~\ref{sec:eval} for methodology.}
	\label{fig:radar}
\end{figure}

%% file: case.tex
\section{A case for a new video store}
\label{sec:case}

We set to design a video store that automatically creates and manages video formats in order to satisfy algorithmic video consumers with high resource efficiency.

\subsection{The Configuration Problem}
\label{sec:case:req}

The store must determine a global set of video formats as follows.
\textbf{Storage format:}
the system may save one ingested stream in multiple versions, each characterized by a fidelity option $f$ and a coding option $c$.
We refer to \FS{}$\langle f,c \rangle$ as a \fs{}.
\textbf{Consumption format:}
the system supplies raw frame sequences to different operators running at a variety of accuracy levels, i.e., \sink{}s.
The format of each raw frame sequence is characterized by a fidelity option $f$.
We refer to \FC{}$\langle f \rangle$ as a \fc{}. 


We refer to the global set of video formats as the store's \textit{configuration} of video formats. 

\Paragraph{Configuration requirements}
These formats should jointly meet the following requirements: 

\Paragraph{R1. Satisfiable fidelity} 
To supply frames in a \fc{} \FC{}$\langle f \rangle$, the system must retrieve video in \fs{} \FS{}$\langle f',c \rangle$, where $f'$ is richer than or the same as $f$.

\input{fig-decoding}

\Paragraph{R2. Adequate retrieving speed}
Video retrieval should not slow down frame consumption. 
Figure~\ref{fig:outpace} show two cases where the slowdown happens. 
a) 
For fast operators sparsely sampling video data, 
decoding may not be fast enough
if the on-disk video is in the original format as it is ingested (e.g., 720p at 30 fps as from a surveillance camera). 
These \sink{}s benefit from \fs{}s that are cheaper to decode, e.g., with reduced fidelity. 
b) For some operators quickly scanning frames looking for simple visual features, 
even the \fs{} that is cheapest to decode 
(i.e., f' is the same as f; cheapest coding option) 
is too slow. 
These \sink{}s benefit from retrieving raw frames from disks. 
\Paragraph{R3. Consolidating \fs{}s}
Each stored video version incurs ingestion and storage costs. 
The system should exploit a key opportunity: 
creating one \fs{} for supplying data to multiple \sink{}s, 
as long as satisfiable fidelity and adequate retrieving speed are ensured. 


\Paragraph{R4. Operating under resource budgets}
The store should keep the space cost by all videos under the available disk space. 
It should keep the ingestion cost for creating all video versions under the system's transcoding bandwidth. 


\input{bkgnd-inadequacy}

%% file: fig-decoding.tex


%

\begin{figure}[t!]
	\centering
	\includegraphics[width=0.47\textwidth{}]{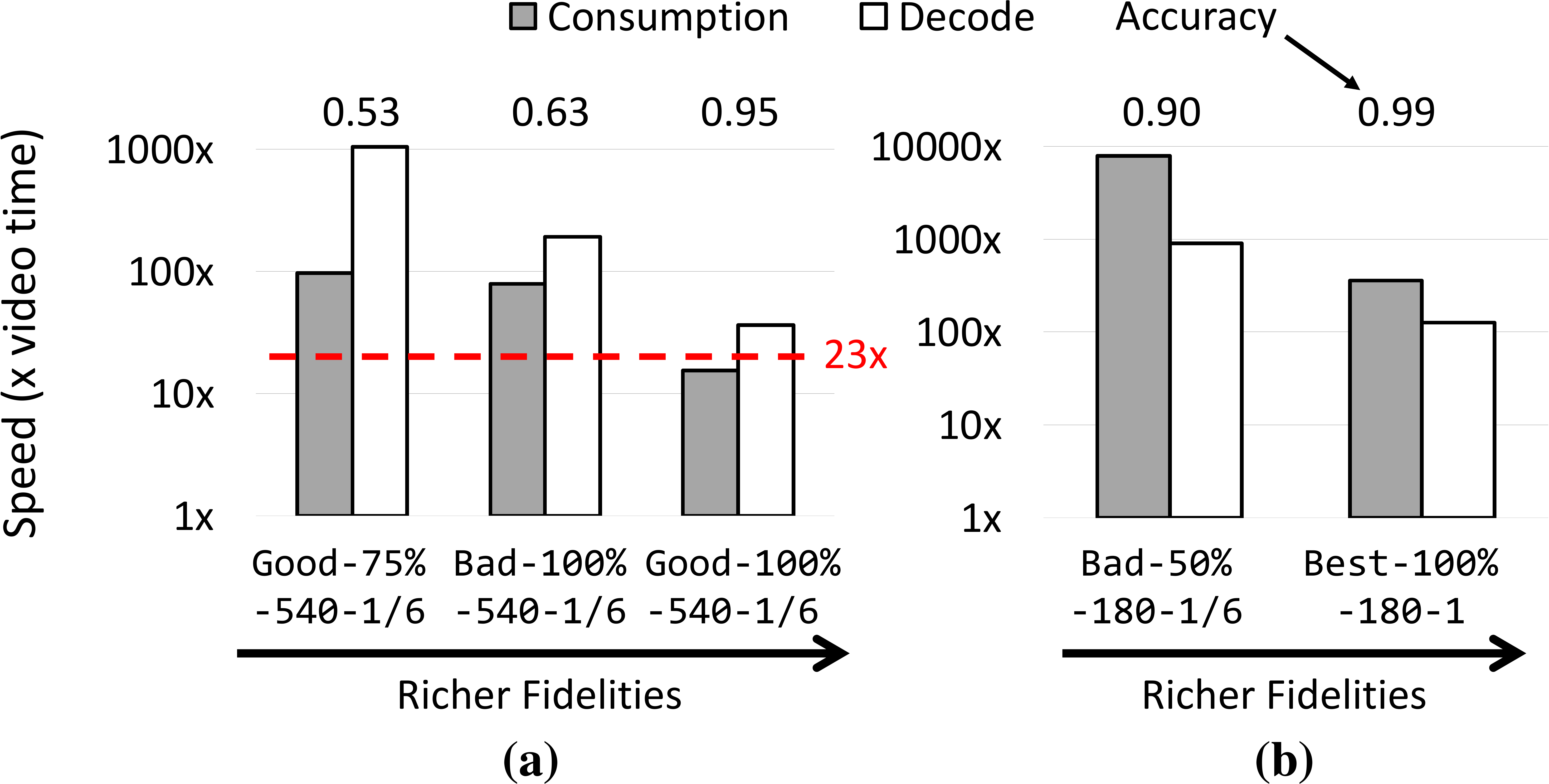} 
	\caption{
		\textbf{Video retrieval \textit{could} bottleneck consumption}. 
		This is exemplified by the decoding speed vs. consumption speed comparisons for two different operators. 
		\textbf{(a)} 
		Operator: License.
		Consumption can be faster than decoding (speed shown as the dashed line), if the on-disk video is stored with the richest fidelity as ingested. 
		Yet, consumption is still slower than decoding video of the same fidelity (white columns). 
		\textbf{(b)} Operator: Motion.
		Consumption is faster than decoding, even if the on-disk video is of the same fidelity as consumed. 
		Operator accuracy annotated on the top. 
		See Section ~\ref{sec:eval} for test hardware.		
	}
	\label{fig:outpace}
\end{figure}

%% file: bkgnd-inadequacy.tex

\subsection{Inadequacy of existing video stores}
Computer vision research typically assumes all the input data present in memory as raw frames, which 
does not hold for retrospective analytics over large videos: 
a server with 100 GB DRAM holds no more than two hours of raw frames even in low fidelity (e.g., 360p at 30 fps). 
Most video stores choose video formats in ad hoc manners \textit{without optimizing for analytics}~\cite{rollingdb}. 
On one extreme, many save videos in one unified format (e.g., the richest fidelity expected by all operators). 
This minimizes storage and ingestion costs while incurring high retrieval cost.
As a result, data retrieval may bottleneck operators.
On the other extreme, one may incarnate all the \fs{}s with the fidelity exactly matching \sink{} expectations.
This misses the opportunities for consolidating \fs{}s and will lead to excessive storage and ingestion costs.
We will evaluate these two alternatives in \sect{eval}. 

\input{layered-coding}

%% file: layered-coding.tex

\Paragraph{Layered encoding cannot simplify the problem}
Layered encoding promises space efficiency: it stores one video's multiple fidelity options as complementary layers~\cite{seufert2015survey}.
%
However, layered encoding has important caveats. 
i) Each additional layer has non-trivial storage overhead (sometimes 40\%--100\%)~\cite{kreuzberger15mmsys}
which may result in storage space \textit{waste} compared to consolidated \fs{}s.  
ii) Decoding is complex and slow, due to the combination of layers and
random disk access in reading the layers. 
iii) 
Invented two decades ago, 
its adoption and coding performance are yet to be seen. 
Even if it is eventually adopted and proven desirable, 
it would make the configuration more complex.

%% file: istream.tex

\section{The \sys{} Design}
\label{sec:design}
\label{sec:overview}



\subsection{Overview}
\label{sec:design:overview}

\input{tab-oplist}  
\sys{} runs on one or over a few commodity servers. 
It depends on existing query executors, e.g., OpenALPR, and a pre-defined library of operators. 
From the executor, \sys{} expects an interface for executing individual operators for profiling, and a manifest specifying a set of option accuracies for each operator.
Table~\ref{tab:oplist} listed 9 operators that are supported by the current \sys{} prototype.
\sys{} tracks the whole set of <operator, accuracy> tuples as \textit{\sink{}s}. 

\Paragraph{Operation}
During operation, \sys{} periodically updates its video format configuration.
For each ingested video stream,
it periodically profiles operators and encoding/decoding, e.g., on a 10-second clip per hour.
\sys{} splits and saves video footage in segments, which are 8-second video clips in our implementation. 
\sys{} retrieves or deletes each segment independently. 


\Paragraph{Challenges}
The major challenges are in configuration.
i) Exhaustive search is infeasible.
A configuration consists of a set of \fc{}s from the 4D space $\mathbb{F}$
and a set of \fs{}s from the 7D space $\mathbb{F} \times \mathbb{C}$.
In our prototype, the total possible global configurations are $24^{15150}$.
ii) 
Exhaustive profiling is expensive, as will be discussed in \sect{design:fc}.
iii) Optimizing for multiple resource types further complicates the problem. 

These challenges were unaddressed. 
Some video query engines seek to ease configuration and profiling (challenge i and ii), 
but are limited to a few knobs~\cite{videostorm,chameleon}. 
For the extensive set of knobs we consider, some of their assumptions, e.g., knob independence, do not hold.
They optimize for one resource type -- GPU cycles for queries, without accounting for other critical resources, e.g., storage  (challenge 3). 


\input{fig-mapping}

\Paragraph{Mechanism overview -- backward derivation}
\sys{} derives the configuration \textit{backwards}, 
in the direction opposite to the video data flow
-- from sinks, to retrieval, and to ingestion/storage. 
This is shown in Figure~\ref{fig:mapping} \circled{1}--\circled{3}.
In this backward derivation, \sys{} optimizes for different resources in a progressive manner. 

\noindent
\circled{1} \sect{design:fc}:
From all given \sink{}s, \sys{} derives video \fc{}s.
Each \sink{} consumes, i.e., subscribes to, a specific \fc{}.
In this step, \sys{} optimizes data consumption speed. 


\noindent
\circled{2} \sect{design:fs}:
From the \fc{}s, 
\sys{} derives \fs{}s.
Each \fc{} subscribes to one \fs{} (along the reversed directions of dashed arrows in Figure~\ref{fig:mapping}). 
The chosen \fs{}s ensure i) satisfiable fidelity: 
a \fs{} \FS{} has richer fidelity than any of its downstream \fc{}s (\FC{}s);
ii) adequate retrieval speed: 
the retrieval speed of \FS{} should exceed the speed of any downstream \sink{} (following the dashed arrows in Figure~\ref{fig:mapping}). 
In this step, \sys{} optimizes for storage cost and keeps ingestion cost under budget. 

\noindent
\circled{3} \sect{design:erosion}:
From all the derived \fs{}s, 
\sys{} derives a data erosion plan, gradually deleting aging video. 
In this step, \sys{} reduces storage cost to be under budget. 

\Paragraph{Limitations} 
\begin{enumerateinline}
\item 
\sys{} treats individual \sink{}s as independent without considering their dependencies in query cascades.
If consumer A always precedes B in all possible cascades, the speed of A and B should be considered in conjunction. 
This requires \sys{} to model all possible cascades, which we consider as future work. 
\item 
\sys{} does not manage algorithmic knobs internal to operators~\cite{chameleon,videostorm}; 
doing so would allow new, useful trade-offs for consumption but not for ingestion, storage, or retrieval.
\end{enumerateinline}

\subsection{Configuring \fc{}s} 
\label{sec:design:fc}

\Paragraph{Objective}
For each \sink{} $\langle op, accuracy \rangle$, 
the system decides a \fc{} $\langle f_0 \rangle$ for the frames supplied to $op$. 
By consuming the frames, $op$ should achieve the target accuracy while consuming data at the highest speed, i.e., with a minimum consumption cost. 

The primary overhead comes from operator profiling. 
Recall the relation 
$ f \rightarrow \langle consumption \: cost, accuracy \rangle$
has to be profiled per operator regularly.
For each profiling, the store prepares sample frames in fidelity f, runs an operator over them, and measures the accuracy and consumption speed.
If the store profiles all the operators over all the fidelity options, 
the total number of required profiling runs, even for our small library of 9 operators is 2.7K.
The total profiling time will be long, as we will show in the evaluation. 

%
 
\Paragraph{Key ideas} 
\sys{} explores the fidelity space efficiently and only profiles a small subset of fidelity options. 
It works based on two key observations.  
%
\textbf{O1. Monotonic impacts} 
Increase in any fidelity knob leads to non-decreasing change in consumption cost and operator accuracy -- 
richer fidelity will neither reduce cost nor  accuracy. 
This is exemplified in Figure~\ref{fig:motiv} showing the impact of changes to individual knobs. 
\textbf{O2. Image quality does not impact consumption cost.} 
Unlike other fidelity knobs controlling data quantity, image quality often does not affect operator workload and thus the consumption cost, as shown in Figure~\ref{fig:motiv}(b).


We next sketch our algorithm deciding the \fc{} for the \sink{} $\langle op,$ accuracy-t$\rangle$: 
the algorithm aims finding $f_0$ that leads to accuracy higher than accuracy-t (i.e., adequate accuracy) with the lowest consumption cost. 

\Paragraph{Partitioning the 4D space} 
i) Given that image quality does not impact consumption cost
(O2), \sys{} starts by temporarily fixing the image quality knob at its highest value.
ii) 
In the remaining 3D space (crop factor $\times$ resolution $\times$ sampling rate), \sys{} searches for fidelity $f_0'$ that leads to adequate accuracy and the lowest consumption cost. 
iii) 
As shown in Figure~\ref{fig:2d-search}, \sys{} partitions the 3D space into a set of 2D spaces for search.
To minimize the number of 2D spaces under search, 
\sys{} partitions along the shortest dimension, chosen as the crop factor which often has few possible values (3 in our implementation). 
iv) 
The fidelity $f_0'$ found from the 3D space already leads to adequate accuracy with the lowest consumption cost.
While lowering the image quality of $f_0'$ does not reduce the consumption cost, 
\sys{} still keeps doing so until the resultant accuracy becomes the \textit{minimum} adequacy. It then selects the knob values as $f_0$.  
This reduces other costs (e.g., storage) opportunistically.



\input{fig-2d-search}

\Paragraph{Efficient exploration of a 2D space}
The kernel of the above algorithm is to search each 2D space (resolution $\times$ sampling rate), as illustrated in Figure~\ref{fig:2d-search}. 
In each 2D space, \sys{} looks for an \textit{accuracy boundary}.
As shown as shaded cells in the figure, the accuracy boundary splits the space into two regions:
all points on the left have \textit{inadequate} accuracies, while all on the right have \textit{adequate} accuracies.
To identify the boundary, \sys{} leverages the fact that accuracy is monotonic along each dimension (O1). 
As shown in Figure~\ref{fig:2d-search}, it starts from the top-right point 
and explores to the bottom and to the left.
\sys{} only profiles the fidelity options on the boundary. 
It dismisses points on the left due to inadequate accuracies.
It dismisses any point X on the right because X has fidelity richer than one boundary point Y; therefore, X incurs no less consumption cost than Y. 

This exploration is inspired by a well known algorithm in searching in a monotone 2D array~\cite{cheng08dm}. 
However, our problem is different: $f_0'$ has to offer both adequate accuracy and lowest consumption cost.
Therefore, \sys{} has to explore the entire accuracy boundary:
its cannot stop at the point where the minimum accuracy is found, which may not result in the lowest consumption cost. 




\Paragraph{Cost \& further optimization}
Each consumer requires profiling runs as many as 
$O((N_{sample} + N_{res}) * N_{crop} + N_{quality})$, where $N_x$ is the number of values for knob $x$.
This is much lower than exhaustive search which requires 
($N_{sample} N_{res} N_{crop} N_{quality}$) runs. 
Furthermore, in profiling for the same operator's different accuracies,
\sys{} memoizes profiling results. 
Our evaluation (Section ~\ref{sec:eval}) will show that profiling \textit{all} accuracies of one operator is still cheaper than exhaustively profiling the operator over the entire fidelity space. 


\Paragraph{What if a higher dimensional fidelity space?}
The above algorithm searches in the 4D space of the four fidelity knobs we consider. 
One may consider additional fidelity knobs (e.g., color channel). 
To search in such a space, we expect partitioning the space along shorter dimensions to still be helpful; 
furthermore, the exploration of 2D space can be generalized for higher dimensional spaces, by retrofitting selection in a high-dimensional monotonic array~\cite{cheng08dm,linial85ja}.

\input{design-fs}

\input{design-erosion}

%% file: tab-oplist.tex

\begin{table}[]
	\footnotesize
	\centering
	\label{tab:definitions}
	\vspace*{-1em}
	\begin{tabular}{@{}l@{$\;\;$}p{2.65in}@{}}
		\toprule[0.7pt]
		\textsf{\textrm{Op}} &
		\textsf{\textrm{Description}} \\ \toprule
	Diff & Difference detector that detects frame differences ~\cite{noscope}\\ 
	S-NN & Specialized NN to detect a specific object ~\cite{noscope}\\ 
	NN & Generic Neural Networks, e.g., YOLO ~\cite{yolo} \\ 
	Motion & Motion detector using background subtraction ~\cite{openalpr} \\ 
	License & License plate detector ~\cite{openalpr}       \\ 
	OCR & Optical character recognition ~\cite{openalpr} \\ 
	Opflow & Optical flows for tracking object movements ~\cite{opflow}  \\ 
	Color & Detector for contents of a specific color ~\cite{blazeit} \\  
	Contour & Detector for contour boundaries ~\cite{contour}\\
		\midrule[0.7pt]
	\end{tabular}
	\caption{The library of operators in the current \sys{}.}
	\label{tab:oplist}
\end{table}

%% file: fig-mapping.tex

\begin{figure}[t!]
\centering
\includegraphics[width=0.48\textwidth{}]{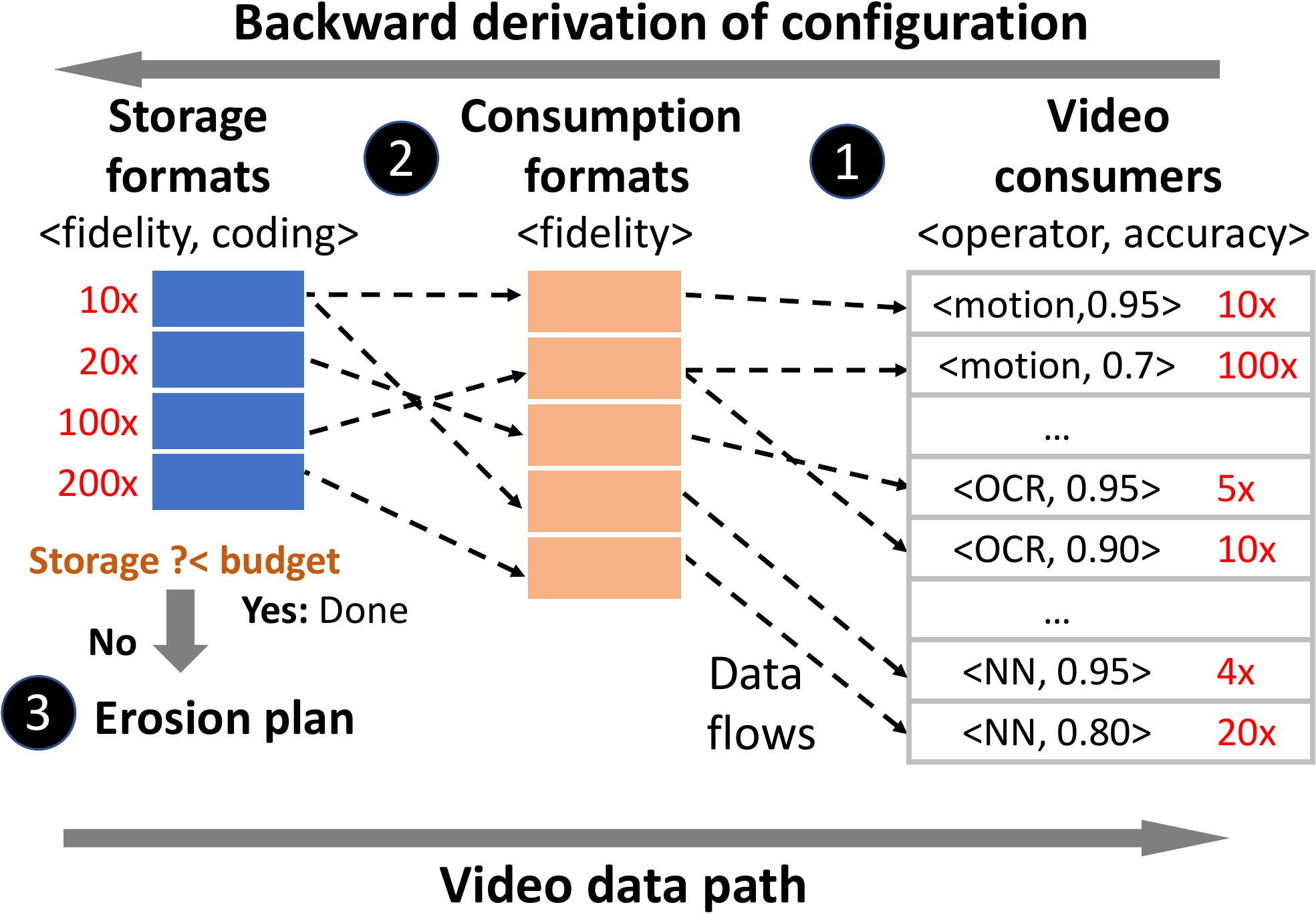} 
\caption{\sys{} derives the configuration of video formats. 
Example consumption/retrieval speed is shown.}
\label{fig:mapping}
\end{figure}

%% file: fig-2d-search.tex

\begin{figure}[t!]
	\centering
	\includegraphics[width=0.45\textwidth{}]{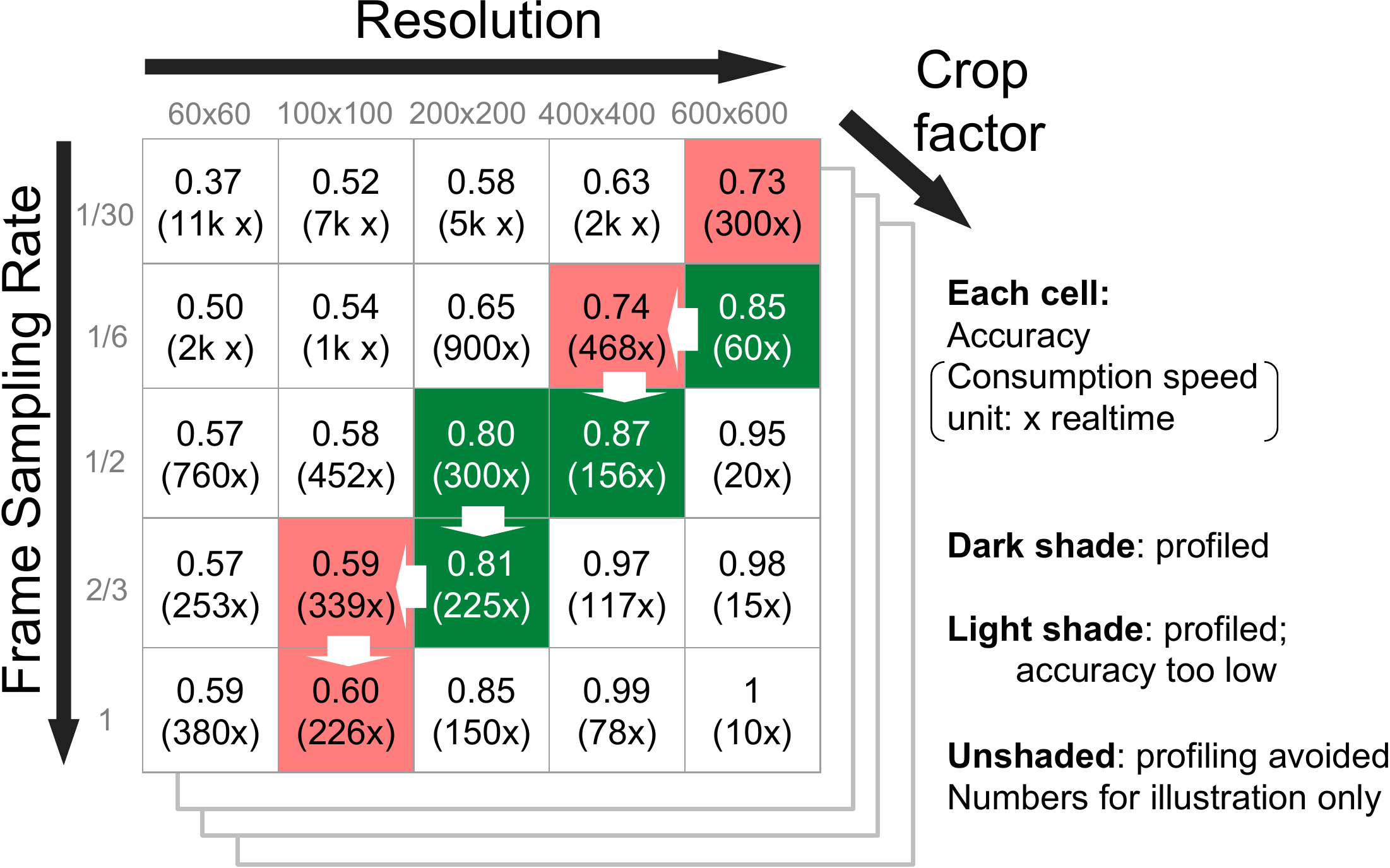} 
	\vspace{-10pt}		
	\caption{
	Search in a set of 2D spaces for a fidelity option with accuracy $\geq$ 0.8 and max consumption speed (i.e., min consumption cost).  
	}
	\label{fig:2d-search}
\end{figure}

%% file: design-fs.tex

\subsection{Configuring \fs{}s} 
\label{sec:design:fs}

\Paragraph{Objective}
For the chosen \fc{}s and their downstream \sink{}s,
\sys{} determines the \fs{}s with satisfiable fidelity and adequate retrieval speed.

\Paragraph{Enumeration is unaffordable}
One may consider enumerating all possible ways to partition the set of consumption formats (\FC{}s), and determining a common \fs{} for each subset of \FC{}s.
This enumeration is very expensive: 
the number of possible ways to partition a \FC{} set is 4$\times10^{6}$ for 12 \FC{}s, and 4$\times10^{17}$ for the 24 \FC{}s in our implementation~\cite{bell1,bell2}.

\input{fig-coalesce}

\Paragraph{Algorithm sketch}
\sys{} coalesces the set of \fs{}s iteratively.
Show on the right side of Figure~\ref{fig:coalesce}, \sys{} starts from a full set of \fs{}s (\FS{}s), each catering to a \FC{} with identical fidelity. 
In addition, \sys{} creates a \textit{golden} storage format \sfg{}<fg,cg>: 
fg is the knob-wise maximum fidelity of all \FC{}s; 
cg is the slowest coding option incurring the lowest storage cost. 
The golden \FS{} is vital to data erosion to be discussed in \sect{design:erosion}. 
All these \FS{}s participate in coalescing. 



\Paragraph{How to coalesce a pair?}
\sys{} runs multiple rounds of pairwise coalescing.
To coalesce \FS{}0$\langle$f0,c0$\rangle$ and \FS{}1$\langle$f1,c1$\rangle$ into \FS{}2$\langle$f2,c2$\rangle$, \sys{} picks f2 to be the knob-wise maximum of f0 and f1 for satisfiable fidelity. 
Such coalescing impacts resource costs in three ways. 
i) It reduces the ingestion cost as the video versions are fewer. 
ii) It may increase the retrieval cost, as \FS{}2 with richer fidelity tends to be slower to decode than \FS{}0/\FS{}1. 
\sys{} therefore picks a cheaper coding option (c2) for \FS{}2, so that decoding \FS{}2 is fast enough for all previous \sink{}s of \FS{}0/\FS{}1. 
Even if the cheapest coding option is not fast enough, \sys{} bypasses coding and stores raw frames for \FS{}2. 
iii) The cheaper coding in turn may increase storage cost. 

\Paragraph{How to select the coalescing pair?}
Recall that the goal of coalescing is to bring the ingestion cost under the budget. 
We explore two alternative approaches. 
\begin{myitemize}
\item 
\textbf{\textit{Distance-based selection.}}
As this is seemingly a hierarchical clustering problem, one may coalesce formats based on their similarity, for which a common metric is Euclidean distance. 
To do so, one may normalize the values of each knob and coalesce the pair of two formats that have the shortest distance among all the remaining pairs. 

\item 
\textbf{\textit{Heuristic-based selection.}}
We use the following heuristics: first harvesting ``free'' coalescing opportunities, and then coalescing at the expense of storage.
Figure~\ref{fig:coalesce} illustrates this process. 
From the right to the left, \sys{} first picks up the pairs that can be coalesced to reduce ingestion cost \textit{without} increasing storage cost. 
Once \sys{} finds out coalescing any remaining pair would \textit{increase} storage cost, 
\sys{} checks if the current total ingestion cost is under budget. 
If not, \sys{} attempts to pick up cheaper coding options and continues to coalesce at the expense of increased storage cost, until the ingestion cost drops below the budget.
\end{myitemize}

%

\Paragraph{Overhead analysis}
The primary overhead comes from profiling. 
Being simple, distance-based selection incurs lower overhead: for each round, it only profiles the ingestion cost of the coalesced \FS{}.
Given that \sys{} coalesces at most $N$ rounds ($N$ being the number of \FC{}s), the total profiling runs are min(O($N$), $|\mathbb{F} \times \mathbb{C}|$). 

By comparison, heuristic-based selection tests all possible pairs among the remaining \FS{}s in each round; 
for each pair, \sys{} profiles a video sample with the would-be coalesced \FS{}, measuring decoding speed and the video sample size. 
The total profiling runs are min(O($N^3$), $|\mathbb{F} \times \mathbb{C}|$). 
In our implementation, $N$ is 24 and $|\mathbb{F} \times \mathbb{C}|$ is 15K. 
Fortunately, by memoizing the previously profiled \FS{}s in the same configuration process, \sys{} can significantly reduce the profiling runs, as we will evaluate in \sect{eval}. 
Furthermore, \sect{eval} will show that heuristic-based selection produces much more compact \FS{}s.

%% file: fig-coalesce.tex
\begin{figure}[t!]
	\centering
	\includegraphics[width=0.45\textwidth{}]{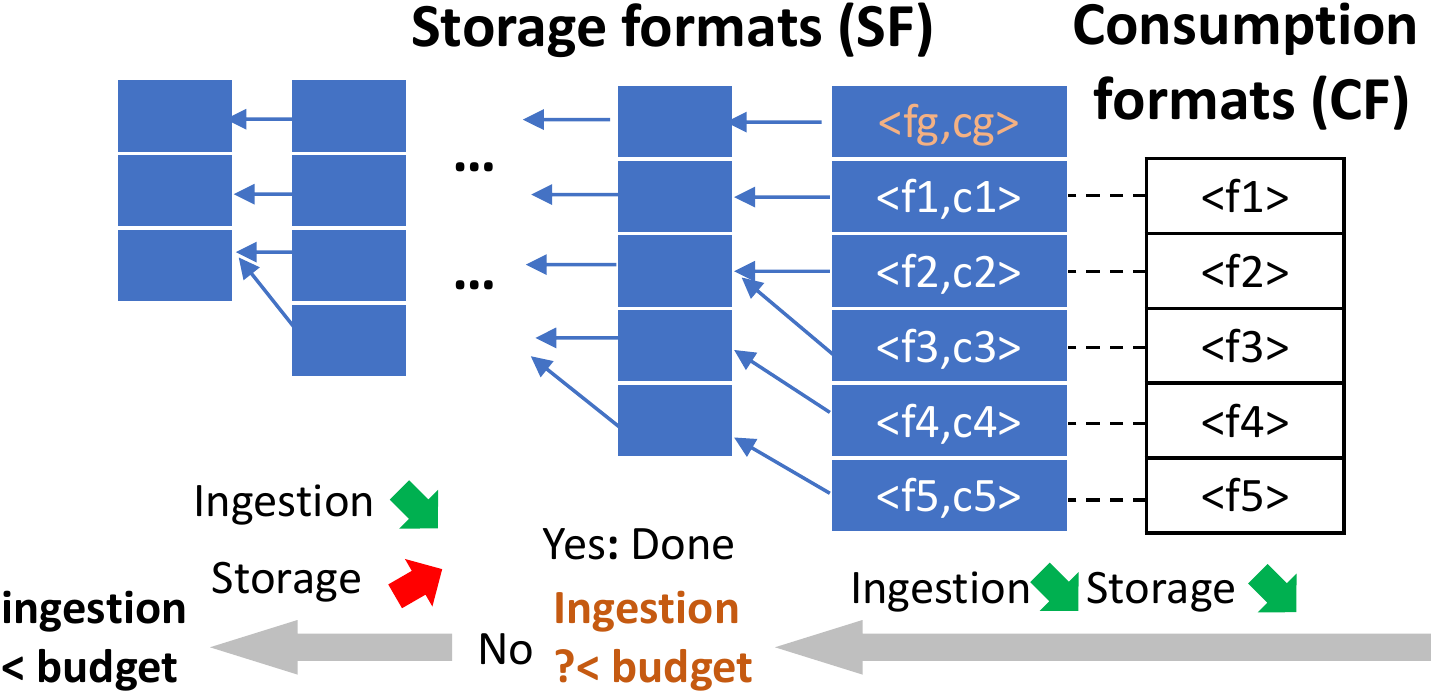} 
	\caption{
	Iterative coalescing of \fs{}s.}
	\label{fig:coalesce}
\end{figure}

%% file: design-erosion.tex

\subsection{Planning Age-based Data Erosion}
\label{sec:design:erosion}


\Paragraph{Objective}
In previous steps, 
\sys{} plans multiple \fs{}s of the same content catering to a wide range of consumers. 
In the last step, \sys{} reduces the total space cost to be below the system budget. 

Our insight is as follows.
As video content ages, 
the system may slowly give up some of the formats, 
freeing space by relaxing the requirement for adequate retrieving speed on aged data (\sect{case}, R2). 
We made the following choices. 
i) 
\textbf{Gracefully degrading consumption speed.} 
\sys{} controls the rate of decay in speed instead of in storage space, as operator speed is directly perceived by users. 
ii) 
\textbf{Low aging cost.} 
\sys{} avoids transcoding aged videos which compete for encoder with ingestion. 
It hence creates no new \fs{}s for aging. 
iii) 
\textbf{Never breaks fidelity satisfiability.} 
\sys{} identifies some video versions as fallback data sources for others, ensuring all \sink{}s to achieve their desired accuracies as long as the videos are still in lifespan.


\input{fig-erosion}


\Paragraph{Data erosion plan}
\sys{} plans erosion at the granularity of video ages.
Recall that \sys{} saves video as segments on disks (each segment contains 8-second video in our implementation). 
As shown in Figure~\ref{fig:erosion},
for each age (e.g., per day) and for each \fs{}, the plan dictates the percentage of deleted segments, which accumulate over ages. 

\Paragraph{How to identify fallback video formats?}
\sys{} organizes all the \fs{}s of one configuration in a tree, where the edges capture \textit{richer-than} relations between the \fs{}s, as shown in Figure~\ref{fig:erosion}. 
Consumers, in an attempt to access any deleted segments of a child format, fall back to the parent format (or even higher-level ancestors). 
Since the parent format offers richer fidelity, the consumers are guaranteed to meet their accuracies; 
yet, the parent's retrieval speed may be inadequate to the \sink{}s 
(e.g., due to costlier decoding), thus decaying the \sink{}s' \textit{effective} speed. 
If a consumer has to consume a fraction $p$ of segments from the parent format, on which the effective speed is only a fraction $\alpha$ of its original speed with no eroded data, 
the consumer's relative speed is defined as the ratio between its decayed speed to its original, given by 
$\alpha/((1-p)\alpha + p)$.
\sys{} never erodes the golden format at the root node;
with its fidelity richer than any other format, the golden format serves as the ultimate fallback for all \sink{}s. 


\Paragraph{How to quantify the overall speed decay?}
Eroding one \fs{} may decay the speeds of multiple \sink{}s to various degrees, 
necessitating a global metric for capturing the overall \sink{} speed. 
Our rationale is for all \sink{}s to fairly experience the speed decay. 
Following the principle of max-min fairness~\cite{maxminfairness},
we therefore define the overall speed as the minimum relative speed of all the \sink{}s.
By this definition, the overall speed $P$ is also relative, in the range of (0,1].
$P$ is 1 when the video content is the youngest and all the versions are intact; 
it reaches the minimum $P_{min}$ when all but the golden format are deleted.


%
%




\Paragraph{How to set overall speed target for each age?}
We follow the power law function, which gives gentle decay rate 
and has been used on time-series data~\cite{summarystore}.
In the function $ P(x) = (1-P_{min})x^{-k} + P_{min} $, $x$ is the video age. 
When $x=1$ (youngest video), $P$ is 1 (the maximum overall speed);
as $x$ grows, $P$ approaches $P_{min}$.
Given a decay factor $k$ (we will show how to find a value below), \sys{} uses the function to set the target overall speed for each age in the video lifespan.

\Paragraph{How to plan data erosion for each age?}
For gentle speed decay, \sys{} always deletes from the \fs{} that would result in the minimum overall speed reduction. 
In the spirit of max-min fairness, 
\sys{} essentially spreads the speed decay evenly among consumers.

\sys{} therefore plans erosion by resembling a fair scheduler~\cite{cfs}. 
For each video age, 
i) \sys{} identifies the \sink{} $Q$ that currently has the lowest relative speed;  
ii) \sys{} examines all \FS{}s in the ``richer-than'' tree, finding the one that has the least impact on the speed of $Q$; 
iii) \sys{} plans to delete a fraction of segments from the found format, so that another \sink{} $R$'s relative speeds drops below $Q$'s.
\sys{} repeats this process until the overall speed drops below the target of this age. 




\Paragraph{Putting it together}
\sys{} generates an erosion plan by testing different values for the decay factor $k$. 
It finds the lowest $k$ (most gentle decay) that brings down the total storage cost accumulated over all video ages under budget.
For each tested $k$, \sys{} generates a tentative plan:
it sets speed targets for each video age based on the power law, plans data erosion for each age, sums up the storage cost across ages, and 
checks if the storage cost falls below the budget. 
As higher $k$ always leads to lower total storage cost, \sys{} uses binary search to quickly find a suitable $k$. 


%% file: fig-erosion.tex

\begin{figure}[t!]
	\centering
	\includegraphics[width=0.45\textwidth{}]{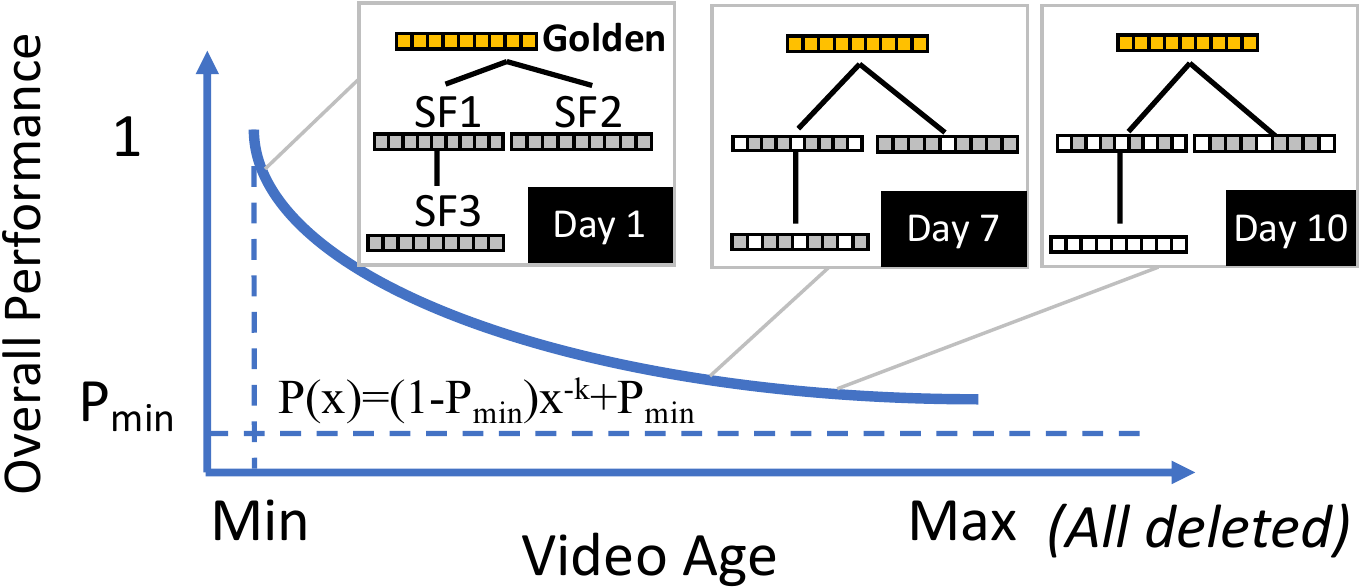} 
	\caption{
	Data erosion decays operator speed and keeps storage cost under budget. Small cells: video segments. 
	}
	\label{fig:erosion}
\end{figure}

%% file: impl.tex

\section{Implementation}
\label{sec:impl}
We built \sys{} in C++ and Python with 10K SLoC. 
Running its configuration engine, \sys{} orchestrates several major components.
%
\textbf{Coding and storage backend:}
\sys{} invokes FFmpeg, a popular software suite for coding tasks.
\sys{}'s ingestion uses the libx264 software encoder; 
it creates one FFmpeg instance to transcode each ingested stream. 
Its retrieval invokes NVIDIA's NVDEC decoder for efficiency. 
\sys{} invokes LMDB, a key-value store ~\cite{lmdb}, as its storage backend. 
\sys{} stores 8-second video segments in LMDB. 
We choose LMDB as it well supports MB-size values. 
\textbf{Ported query engines:}
We ported two query engines to \sys{}. 
We modify both engines so they retrieve data from \sys{} and provide interfaces for \sys{}'s profiling. 
OpenALPR~\cite{openalpr} recognizes vehicle license plates. 
Its operators build on OpenCV and run on CPU. 
To scale up, we create a scheduler that manages multiple OpenALPR contexts and dispatches video segments. 
NoScope~\cite{noscope} is a recent research engine. 
It combines operators that execute at various speeds and invoke deep NN. 
It invokes TensorFlow~\cite{tensorflow} as the NN framework, which runs on GPU. 
\textbf{Operator lib:}
The two query engines provide 6 operators as shown in Figure~\ref{fig:pipeline}.
In particular, S-NN uses a very shallow AlexNet~\cite{alexnet} produced by NoScope's model search and NN uses YOLOv2~\cite{yolo}. 


%% file: eval-sys.tex

\section{Evaluation}
\label{sec:eval}

We answer the following questions in evaluation:
\begin{description}[align=left]
	\item [\S\ref{sec:eval:e2e}] Does \sys{} provide good end-to-end results?
	\item [\S\ref{sec:eval:budget}] Does \sys{} adapt configurations to resource budgets?
	\item [\S\ref{sec:eval:config}] Does \sys{} incur low overhead in configuration?
\end{description}

\subsection{Methodology} 
\label{sec:eval:method}

\Paragraph{Video Datasets}
We carried out our evaluation on six videos, extensively used as benchmarks in prior work~\cite{noscope,focus,blazeit,chameleon}. 
We include videos from both dash cameras (which contain high motion) and surveillance cameras that capture traffic from heavy to light.
The videos are: 
\textit{jackson}, from a surveillance camera at Jackson Town Square;
\textit{miami}, from a surveillance camera at Miami Beach crosswalk;
\textit{tucson}: from a surveillance camera at Tucson 4-th Avenue.
\textit{dashcam}, from a dash camera when driving in a parking lot; 
\textit{park}, from a stationary surveillance camera in a parking lot; 
\textit{airport}, from a surveillance camera at JAC parking lot. 
The ingestion formats of all videos are 720p at 30 fps encoded in H.264. 

\Paragraph{\sys{} setup}
We, as the system admin, declare a set of accuracy levels \{0.95, 0.9, 0.8, 0.7\} for each operator. 
These accuracies are typical in prior work~\cite{chameleon}.
In determining F1 scores for accuracy, we treat as the ground truth when the operator consumes videos in the ingestion format, i.e., highest fidelity. 
In our evaluation, we run the two queries as illustrated in Figure~\ref{fig:pipeline}: 
Query A (Diff + S-NN + NN) and query B (Motion + License + OCR).
In running the queries, we, as the users, select specific accuracy levels for the operators of the query. 
In running queries, we, as the users, specify different accuracy levels for the constituting operators. 
We run query A on the first three videos and B on the remainder, as how these queries are benchmarked in prior work~\cite{openalpr,noscope}.
To derive \fc{}s, \sys{} profiles the two sets of operators on \textit{jackson} and \textit{dashcam}, respectively. 
Each profiled sample is a 10-second clip, a typical length used in prior work~\cite{chameleon}. 
\sys{} derives a unified set of \fs{}s for all operators and videos.

\Paragraph{Hardware environment}
We test \sys{} on a 56-core Xeon E7-4830v4 machine with 260 GB DRAM, 4$\times$1TB 10K RPM SAS 12Gbps HDDs in RAID 5, and a NVIDIA Quadro P6000 GPU. 
By their implementation, the operators from ALPR run on the CPU; we limit them to use up to 40 cores for ensuring the query speed comparable to commodity multi-core servers. 
The operators from NoScope run on the GPU.


\subsection{End-to-end Results}
\label{sec:eval:e2e}

\input{fig-fctable}

\Paragraph{Configuration by \sys{}}
\sys{} automatically configuring video formats based on its profiling. 
Table~\ref{tab:fc-table} shows a snapshot of configuration, including the whole set of \fc{}s (\FC{}s) and \fs{}s (\FS{}s). 
For all the 24 \sink{}s (6 operators at 4 accuracy levels), \sys{} generates 21 unique \FC{}s, as shown in Table~\ref{tab:fc-table}(a). 
The configuration has 109 knobs over all 21 \FC{}s (84 knobs) and 4 \FS{}s (25 knobs), with each knob having up to 10 possible values. 
Manually finding the optimal combination would be infeasible, which warrants \sys{}'s automatic configuration. 
In each column (a specific operator), 
although the knob values \textit{tend} to decrease as accuracy drops, 
the trend is complex and can be non-monotone.
For instance, in column Diff, from F1=0.9 to 0.8, \sys{} advises to decrease sampling rate, while \textit{increase} the resolution and crop factor. 
This reflects the complex impacts of knobs as stated in Section~\ref{sec:bkgnd:knob-impacts}. 
We also note that \sys{} chooses extremely low fidelity for Motion at all accuracies $\leq$ 0.9. 
It suggests
that Motion can benefit from an even larger fidelity space with even cheaper fidelity options. 

From the \FC{}s, \sys{} derives 4 \FS{}s, including one golden format (\sfg{}), as listed in Table~\ref{tab:fc-table}(b).
Note that we, as the system admin, has not yet imposed any budget on ingestion cost.
Therefore, \sys{} by design chooses the set of \FS{}s that minimize the total storage cost (Section ~\ref{sec:design:fs}).
The \FC{} table on the left tags each \FC{} with the \FS{} that each \FC{} subscribes to. 
As shown, the \FC{}s and \FS{}s jointly meet the design requirements R1--R3 in Section~\ref{sec:design:fs}:
each \FS{} has fidelity richer than/equal to what its downstream \FC{}s demand; 
the \FS{}'s retrieval speed is always faster than the downstream's consumption speed. 
Looking closer at the \FS{}s:  
\sfg{} mostly caters to \sink{}s demanding high accuracies but low consumption speeds; 
SF3, stored as low-fidelity raw frames, caters to high-speed \sink{}s demanding low image resolutions; 
between \sfg{} and SF3, SF1 and SF2 fill in the wide gaps of fidelity and costs. 
Again, it is difficult to manually determine such a complementary set of \FS{}s without \sys{}'s configuration.

\Paragraph{Alternative configurations}
We next quantitatively contrast \sys{} with the following alternative configurations: 
\begin{myitemize}
\item 
\textbf{1$\rightarrow$1} stores videos in the golden format (\sfg{} in Table~\ref{tab:fc-table}).
All consumers consume videos in this golden format. 
This resembles a video database oblivious to algorithmic consumers. 
\item
\textbf{1$\rightarrow$N}
stores videos in the golden format \sfg{}.
All consumers consume video in the \FC{}s determined by \sys{}. 
This is equivalent to \sys{} configuring video formats for consumption but not for storage. 
The system, therefore, has to decode the golden format and convert it to various \FC{}s. 
\item 
\textbf{N$\rightarrow$N} 
stores videos in 21 \FS{}s, one for each unique \FC{}. 
This is equivalent to \sys{} giving up its coalescing of \FS{}s. 
\end{myitemize}

\input{eval-e2e-alpr}

\Paragraph{Query speed}
As shown in Figure~\ref{fig:e2e}(a), \sys{} achieves good query speed overall, up to 362$\times$ realtime. 
\sys{}'s speed is incomparable with performance reported for retrospective analytics engines~\cite{noscope,blazeit}:
while \sys{} streams video data (raw/encoded) from disks through decoder to operators,
the latter were tested with all input data preloaded as raw frames in memory. 
\sys{} offers flexible accuracy/cost trade-offs: 
for queries with lower target accuracies, 
\sys{} accelerates query speed by up to 150$\times$.
This is because \sys{} elastically scales down the costs: 
according to the lower accuracies, it switches the operators to \FC{}s that incur lower consumption cost; 
the \FC{}s subscribe to \FS{}s that incur lower retrieval cost. 



Figure~\ref{fig:e2e}(a) also shows the query speed under alternative configurations.
1$\rightarrow$1 achieves the best accuracy (treated as the ground truth) as it consumes video in the full fidelity as ingested. 
However, it cannot exploit accuracy/cost trade-offs, offering a fixed operating point.  
By contrast, \sys{} offers extensive trade-offs and speeds up queries by two orders of magnitude. 

1$\rightarrow$N customizes \fc{}s for \sink{}s while only storing the golden format. 
Although it minimizes the consumption costs for \sink{}s, 
it essentially caps the effective speed of all consumers at the speed of decoding the golden format, which is about 23$\times$ of realtime. 
The bottlenecks are more serious for lower accuracy levels (e.g., 0.8) where many \sink{}s are capable of consuming data as fast as tens of thousand times of realtime, as shown in Table~\ref{tab:fc-table}(a). 
As a result, \sys{} outperforms 1$\rightarrow$N by 3$\times$-16$\times$, demonstrating the necessity of the \FS{} set. 

\Paragraph{Storage cost}
Figure~\ref{fig:e2e}(b) compares the storage costs. 
Among all, 
N$\rightarrow$N incurs the highest costs, because it stores 21 video versions in total. 
For \textit{dashcam}, a video stream with intensive motion which makes video coding less effective, 
the storage cost reaches as high as 2.6 TB/day, filling a 10TB hard drive in four days. 
In comparison, 
\sys{} consolidates the \fs{}s effectively and therefore reduces the storage cost by 2$\times$-5$\times$. 
1$\rightarrow$1 and 1$\rightarrow$N require the lowest storage space as they only save one video version per ingestion stream; 
yet, they suffer from high retrieval cost and low query speed. 

\Paragraph{Ingestion cost}
Figure~\ref{fig:e2e}(c) demonstrates that \sys{} substantially reduces ingestion cost through consolidation of \fs{}s. 
Note that it shows \sys{}'s \textit{worst-case} ingestion cost. 
As stated earlier, in the end-to-end experiment with no ingestion budget imposed,
\sys{}, therefore, reduces the ingestion cost without \textit{any} increase in the storage cost. 
As we will show next, once an ingestion budget is given, \sys{} can keep the ingestion cost much lower than the worst case with only a minor increase in storage cost. 


Overall, on most videos \sys{} requires around 9 cores to ingest one video stream, transcoding it into the 4 \FS{}s in real time (30 fps). 
Ingesting \textit{dashcam} is much more expensive, as the video contains intensive motion.
\sys{}'s cost is 30\%--50\% lower than N$\rightarrow$N, which must transcode each stream to 21 \FS{}s. 
1$\rightarrow$1 and 1$\rightarrow$N incur the lowest ingestion cost as they only transcode the ingestion stream to the golden format, yet at the expense of costly retrieval and slow query speed.





\subsection{Adapting to Resource Budgets}
\label{sec:eval:budget}
\input{fig-ingest-budget}

\Paragraph{Ingestion budget}
\sys{} elastically adapts its configuration with respect to the ingestion budget. 
To impose budget, we, as the system admin, cap the number of CPU cores available to one FFmpeg that transcodes each ingested stream. 
In response to the reduced budget, \sys{} gently trades off storage for ingestion. 
Table~\ref{tab:ingest-budget} shows that as the ingestion budget drops, \sys{} incrementally tunes up the coding speed (i.e., cheaper coding) for individual \FS{}s. 
As a trade-off, the storage cost slowly increases by 17\%.
During this process, the increasingly cheaper coding \textit{overprovisions} the retrieval speed to \sink{}s and therefore will never fail the latter's requirements. 
Note that at this point, the total ingestion output throughput is still less than 3.6 MB/s; even the system ingests 56 streams with its 56 cores concurrently, the required disk throughput 200 MB/s is still far below that of a commodity HDD array (1 GB/s in our platform). 
\input{fig-op-converge}
We also find out that \FS{}s as well as the ingestion cost quickly plateaus as \sys{}'s library includes more operators. 
Figure~\ref{fig:op-converge} shows how the ingestion cost increases as operators are sequentially added, following the order listed in Table~\ref{tab:oplist}, to \sys{}'s library.
The ingestion cost stabilizes as the number of operators exceeds 5, as additional operators share existing \FS{}s. 

\Paragraph{Storage budget}
\input{fig-decay}
\sys{}'s data erosion effectively respects the storage budget with gentle speed decay. 
To test \sys{}'s erosion planning, we, as system admin, set the video lifespan to 10 days; 
we then specify different storage budgets. 
With all 4 \FS{}s listed in Table~\ref{tab:fc-table}(b), 10-day video stream will take up 5 TB of disk space. 
If we specify a budget above 5 TB, \sys{} will determine not to decay ($k$=0), shown as the flat line in Figure~\ref{fig:decay}(a). 
Further reducing the storage budget prompts data erosion.
With a 4 TB budget, \sys{} decays the overall operator speed (defined in Section ~\ref{sec:design:erosion}) following a power law function ($k$=1). 
As we further reduce the budget, \sys{} plans more aggressive decays to respect the budget. 
Figure~\ref{fig:decay}(b) shows how \sys{} erodes individual \fs{}s under a specific budget.
On day 1 (youngest), all 4 \FS{}s are intact. 
As the video ages, \sys{} first deletes segments from \FS{}1 and \FS{}2 that have lower impacts on the overall speed. 
For segments older than 5 days, \sys{} deletes all the data in \FS{}1-3, while keeping the golden format intact (not shown). 

\subsection{Configuration Overhead} 
\label{sec:eval:config}
\sys{} incurs moderate configuration overhead, thanks to our techniques in Section ~\ref{sec:design:fc} and ~\ref{sec:design:fs}. 
Overall, one complete configuration (including all required profiling) takes around 500 seconds, suggesting the store can afford one configuration process in about every 1 hour online.

\input{fig-search}

\Paragraph{Configuring \fc{}s}
Figure~\ref{fig:search} shows the overhead in determining \fc{}s. 
Compared to exhaustive profiling of all fidelity options, 
\sys{} reduces the number of profiling runs by 9$\times$--15$\times$ 
and the total profiling delay by 5$\times$, from 2000 seconds to 400 seconds. 
We notice that the License operator is slow, contributing more than 75\% of total delay, likely due to its CPU-based implementation. 

\label{sec:eval:merge}
\Paragraph{Configuring \fs{}s}
We have validated that \sys{} is able to find resource-efficient \fs{}s as exhaustive enumeration does. 

\textbf{\textit{Heuristic-based selection}}:
We first test heuristic-based selection for producing \FS{}s (Section ~\ref{sec:design:fs}). 
We compare it to exhaustive enumeration, on deriving \FS{}s from the 12 \FC{}s used in query B;
we cannot afford more \FC{}s would which make exhaustive enumeration very slow. 
Both methods result in identical \fs{}s, validating \sys{}'s rationale behind coalescing. 
Yet, \sys{}'s overhead (37 seconds) is 2 orders of magnitude lower than enumeration (5548 seconds).

To derive the \fs{}s from all the 21 unique \fc{}s in our evaluation, 
\sys{} incurs moderate absolute overhead (less than 1 min) too. 
Throughout the 17 rounds of coalescing, it only profiles 475 (3\%) \fs{}s out of all 15K possible ones. 
We observed that its memorization is effective: 
despite 5525 \fs{}s are examined as possible coalescing outcomes, 
92\% of them have been memoized before and thus requires no new profiling. 

\textbf{\textit{Distance-base selection}}:
We then test the other strategy. 
We use Euclidean distance as the similarity metric. 
The configuration takes only 18 seconds, 2$\times$ shorter than the heuristic-based selection mentioned above. 
This is because calculating the distances requires no expensive profiling as heuristic-based selection does. 

\textbf{\textit{Comparison of resultant \FS{}s}}:
The two strategies also derive very different \FS{} sets: 
while the \FS{}s derived by heuristic-based selection is close to optimal as shown above,
the \FS{}s derived by distance-based selection incur 2.2$\times$ higher storage cost.
This is because the latter strategy, while simple, overlooks the fact that different knobs have complex and varying resource impacts (Section ~\ref{sec:bkgnd:knob-impacts}), which cannot be simply normalized across knobs. 

%% file: fig-fctable.tex

\begin{table*}[t!]
	\centering


		\includegraphics[width=1\textwidth]{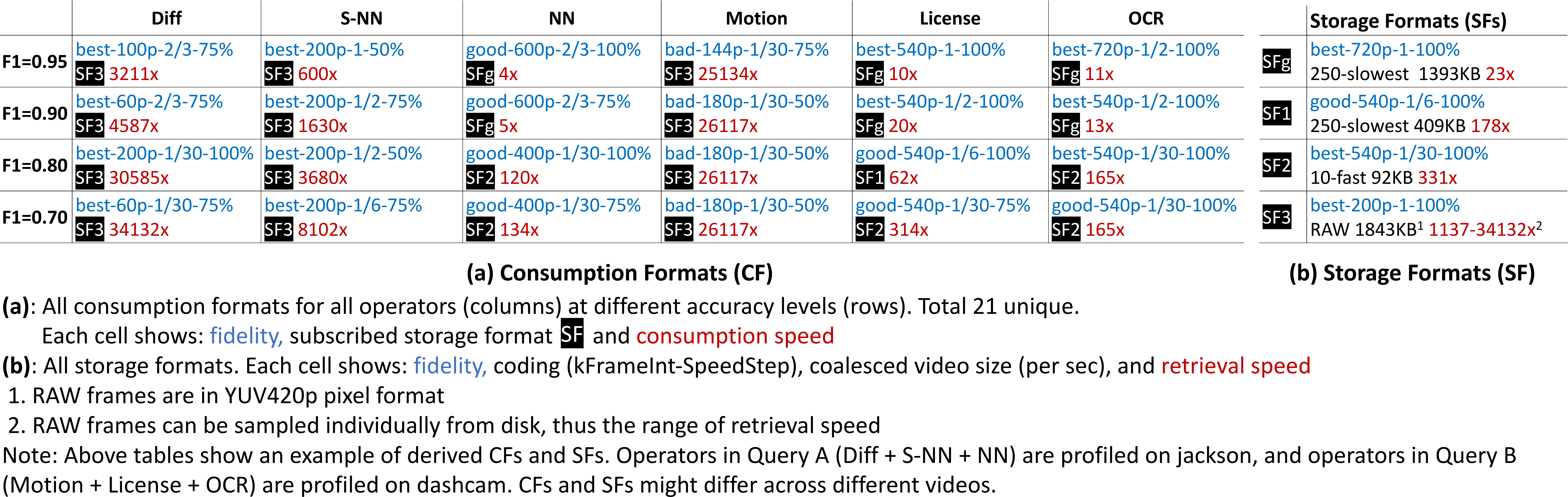}
	
	\caption{A sample configuration of video formats automatically derived by \sys{}.}
	\label{tab:fc-table}
\end{table*}

%% file: eval-e2e-alpr.tex

\begin{figure}
	\centering
	
	
	\begin{minipage}[b]{0.48\textwidth}
			\includegraphics[width=1\textwidth]{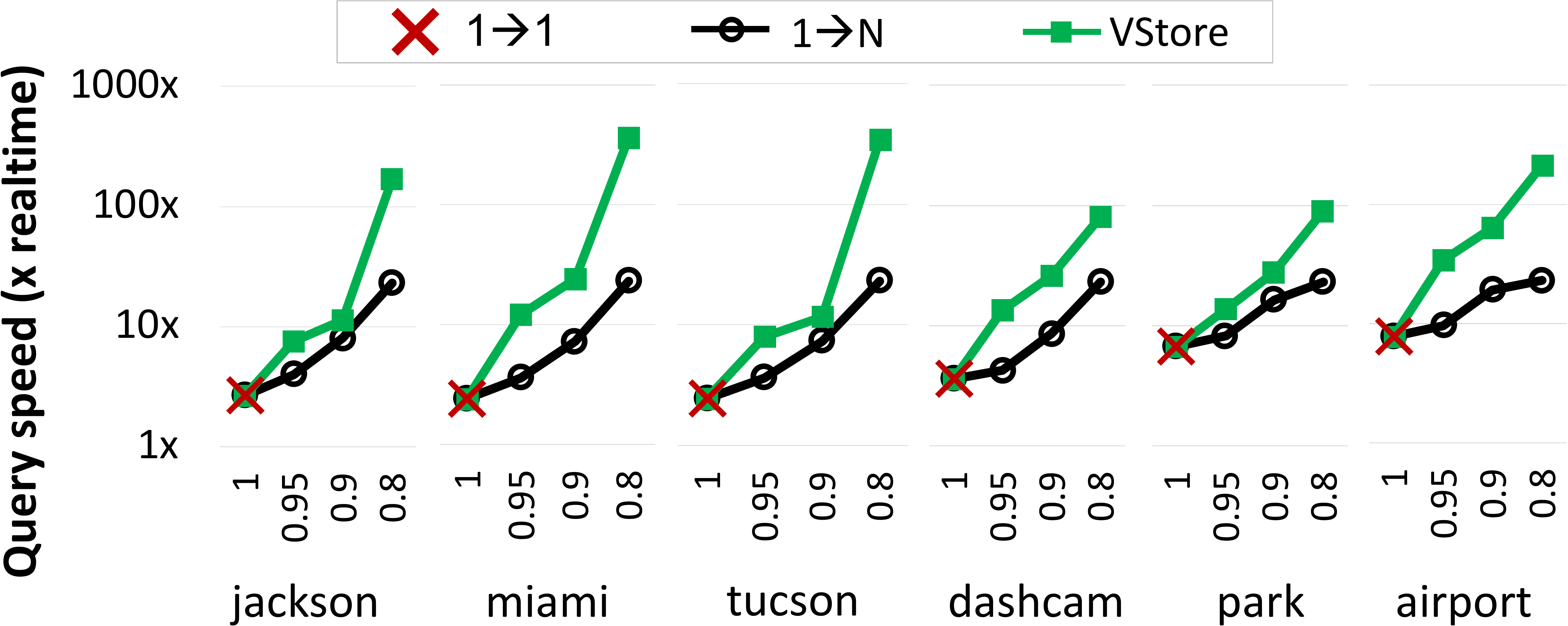}
		\subcaption{Query speeds (y-axis; logarithmic scale) as functions of target operator accuracy (x-axis). 
				Query A on left 3 videos; query B on right 3 videos.
				By avoiding video retrieval bottleneck, \sys{} significantly outperforms others. In this work, we assume the ingestion format is the ground truth (accuracy = 1). Note that N$\rightarrow$N is omitted since the speed is the same as \sys{}.}
				\label{fig:e2e-speed}
	\end{minipage}
	\centering				
	\begin{minipage}[b]{0.48\textwidth}
			\includegraphics[width=1\textwidth]{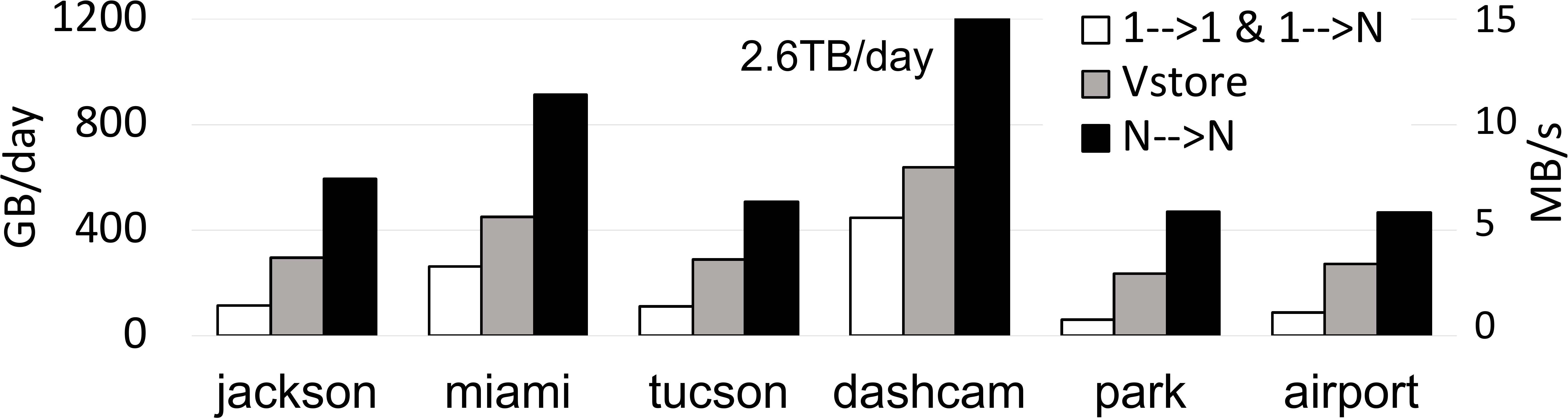}
		\subcaption{Storage cost per video stream, measured as the growth rate of newly stored video size as ingestion goes on. 
		\sys{}'s coalsecing of \FS{}s substantially reduces storage cost. 
		}
				\label{fig:e2e-store}
	\end{minipage}
	\centering				
	\begin{minipage}[b]{0.48\textwidth}
		\includegraphics[width=1\textwidth]{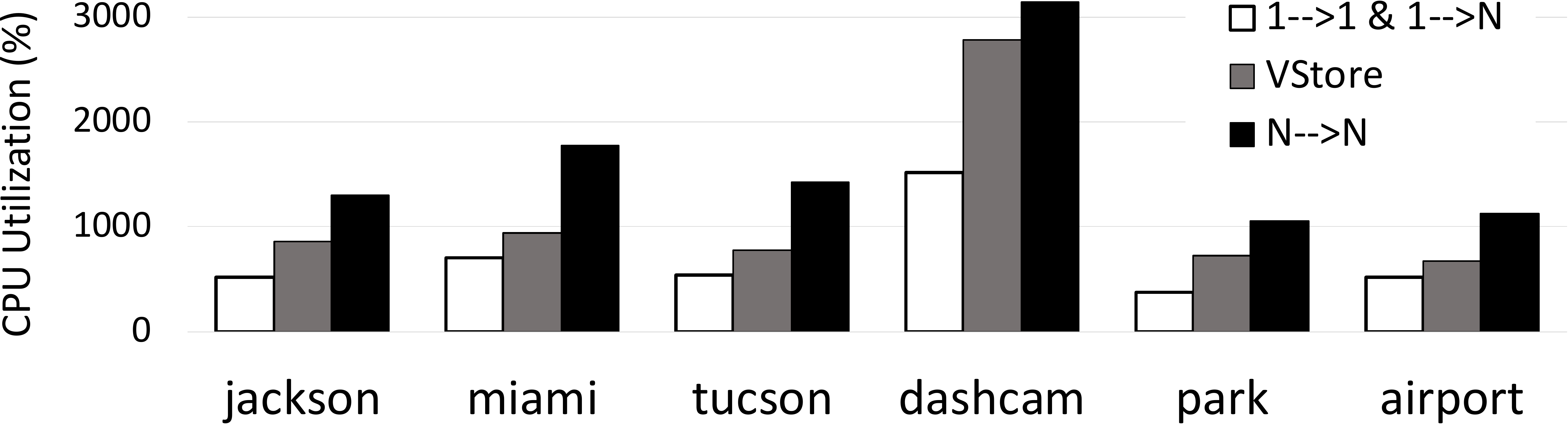}
		\subcaption{Ingestion cost per video stream, as required CPU usage for transcoding the stream into \fs{}s.
		\sys{}'s \FS{} coalescing substantially reduces ingestion cost. 
		Note that this shows \sys{}'s \textit{worst-case} ingestion cost with no ingestion budget specified; 
		see Table~\ref{tab:ingest-budget} for more.}
			\label{fig:e2e-ingest}
	\end{minipage}
	
	\caption{End-to-end result.}
	\label{fig:e2e}
\end{figure}

%% file: fig-ingest-budget.tex

\begin{table}
	\centering
	\includegraphics[width=0.48\textwidth{}]{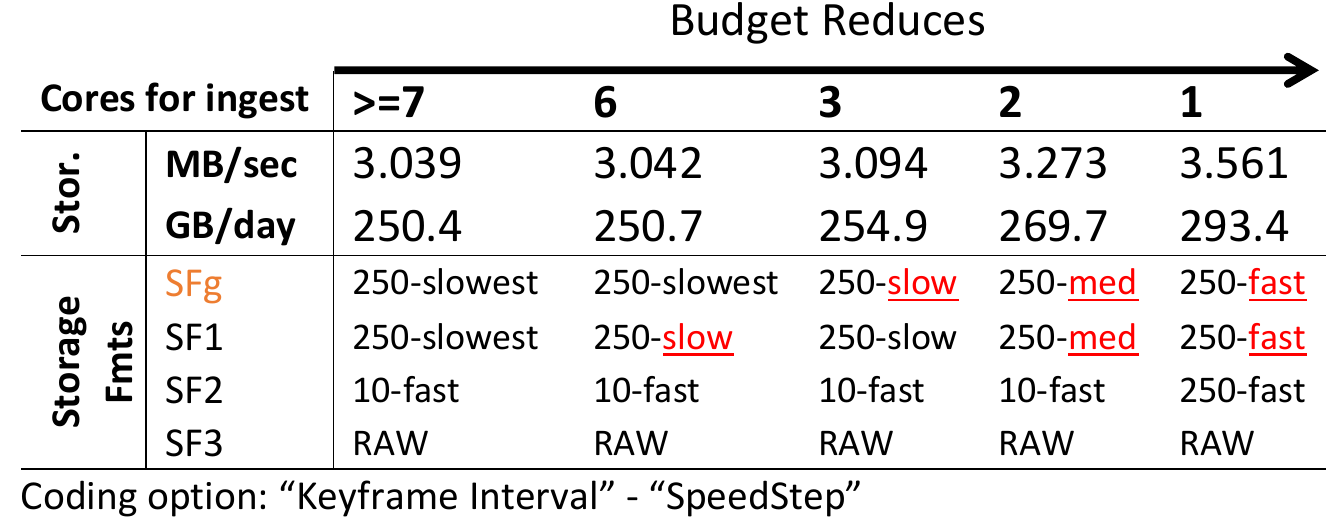} 
	\vspace{-10pt}		
	\caption{In response to ingestion budget drop, \sys{} tunes coding and coalesces formats to stay under the budget with increase in storage cost. Changed knobs shown in \underline{red}.}
	\label{tab:ingest-budget}
\end{table}

%% file: fig-op-converge.tex

\begin{wrapfigure}{r}{0.25\textwidth}
	\centering
	\includegraphics[width=0.24\textwidth{}]{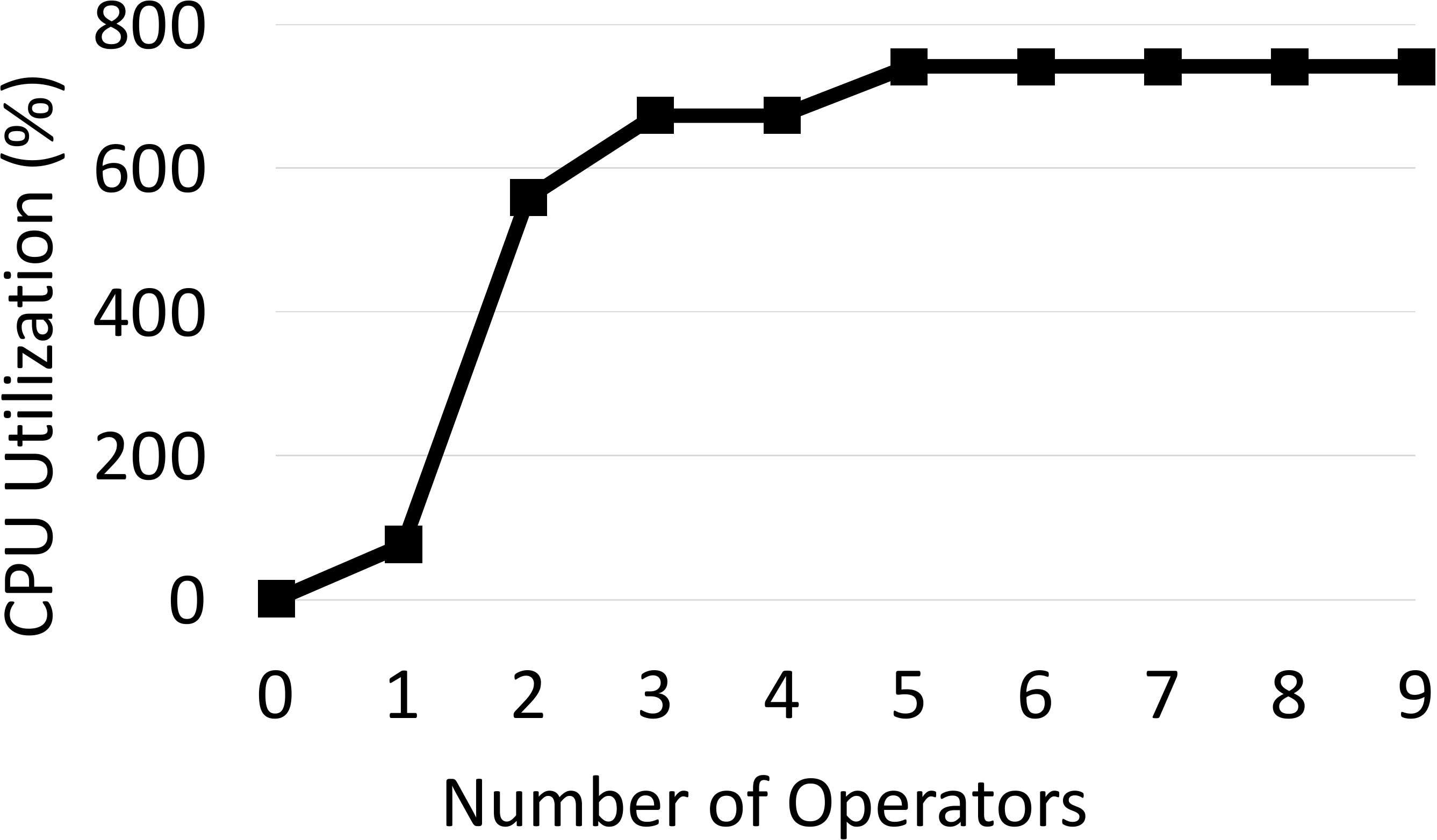} 
	\vspace{-10pt}		
	\caption{Transcoding cost does not scale up with the number of operators. Operator sequence follows Table ~\ref{tab:oplist}.}
	\label{fig:op-converge}
\end{wrapfigure}

%% file: fig-decay.tex
\begin{figure}[t!]
	\centering				
	\begin{minipage}[b]{0.4\textwidth}
		\includegraphics[width=1\textwidth]{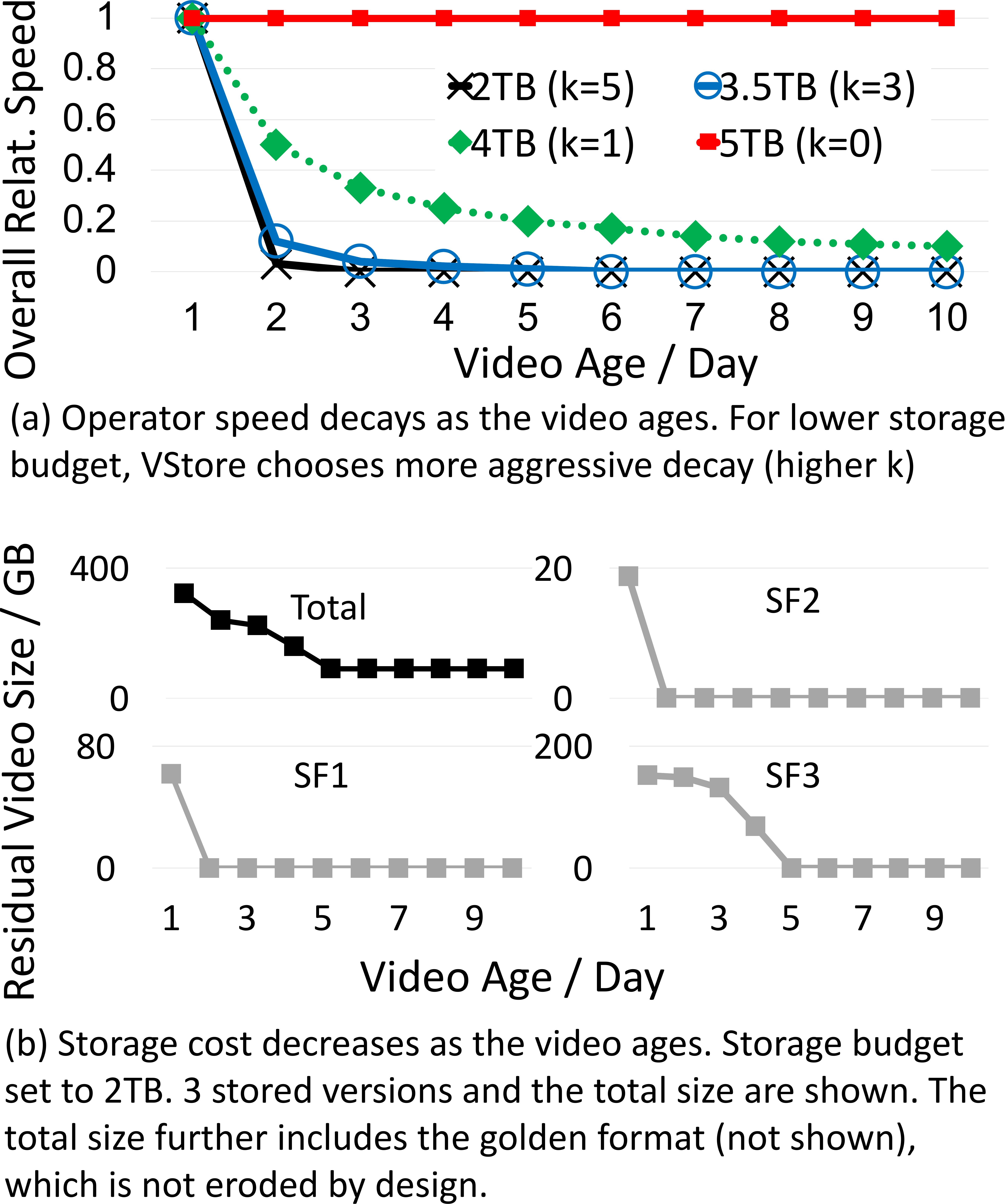}
	\end{minipage}
	~ 
	\caption{Age-based decay in operator speed (a) and reducing storage cost (b) to respect storage budget.}
	\label{fig:decay}
\end{figure}

%% file: fig-search.tex
\begin{figure}[t!]
	\centering
	\includegraphics[width=0.46\textwidth{}]{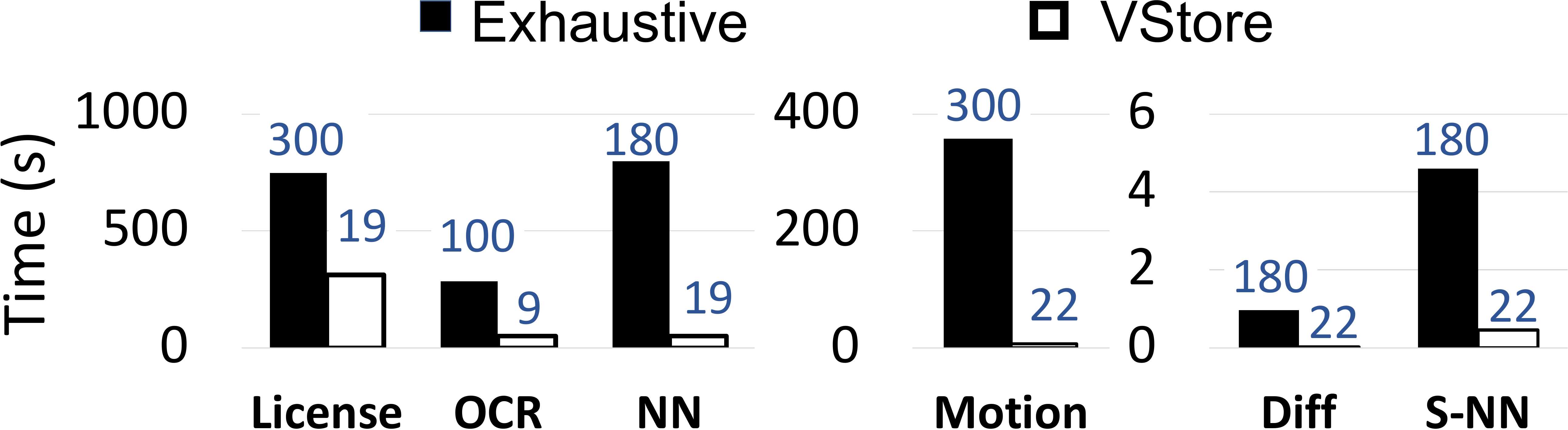} 
	\caption{Time spent on deriving \fc{}s. 	
	Numbers of profiling runs are annotated above columns.
	Each required profiling runs on a 10-second video segment. 
	\sys{} reduces overhead by 5$\times$ in total. 
	}
	\label{fig:search}
\end{figure}


%% file: discussion.tex
\section{Discussion}
\label{sec:discussion}

\Paragraph{Adapting to changes in operators and hardware}
\sys{} works with any possible queries composed by operators/accuracies pre-defined in its library (Section ~\ref{sec:bkgnd:model}). 
If users add a new operator (or a new accuracy level), \sys{} would need to profile the new operator and derive corresponding \FC{}s for it.
If users change the platform hardware (e.g., adding a new GPU), \sys{} would need to re-profile all existing operators. 
Conceptually, this also triggers an update to the \FS{}s. 
Since transcoding existing on-disk videos is expensive, \sys{} only applies the updated \FS{}s to forthcoming videos; 
for existing videos, \sys{} makes each new \FC{} subscribe to the cheapest existing \FS{} with satisfiable fidelity (Section ~\ref{sec:case:req}).
As a result, on existing videos, operators run with designated accuracies, albeit slower than optimal. 
As this portion of videos age and retire, operators run at optimal speed on all videos. 


\Paragraph{Qualitative comparison against Focus~\cite{focus}}
As stated in \sect{bkgnd:model}, 
Focus by design is limited to fixed query pipelines -- object detection consisting of one cheap neural network (NN) and one full NN. 
This contrasts with \sys{} which supports diverse queries. 
Nevertheless, we compare their resource costs on such an object detection pipeline. 

\textbf{\textit{Ingestion cost.}}
\sys{} continuously runs transcoding. 
As already shown in Figure~\ref{fig:op-converge}, the transcoding cost quickly plateaus as the number of operators grows. 
While the current \sys{} prototype runs transcoding on CPU for development ease, low-cost hardware transcoder is pervasive: 
recent work showcases a transcoder farm of \$20 Raspberry Pis, with each device transcoding 4 video streams in real time (720$\times$480 at 30 fps)~\cite{liu16atc, yoon16sec}. 
We, therefore, estimate the hardware cost for each video ingestion to be less than a few dozen dollars.

By comparison, at ingestion time, Focus continuously runs the cheap NN on GPU.
On a high-end GPU (Tesla P100, $\sim$\$4,000), the cheap NN is reported to run at 1.92K fps; 
assuming perfect scalability, this GPU supports up to 60 video streams.
The hardware investment for ingesting each video stream is around \$60, which is 2$\times$-3$\times$ higher than \sys{}. 
If the ingested streams are fewer (e.g., several or a few dozen as typical for a small deployment), the GPU is underutilized, which further increases per-stream investment. 
Running the ingestion on public cloud helps little: 
Amazon EC2's single-GPU instance (P3) costs nearly \$17.5K per year~\cite{amazon-ec2}.



\textbf{\textit{Query cost.}}
At query time, 
\sys{} would run the cheap NN on all frames and
the full NN on the frames selected by the cheap NN. 
By comparison, Focus only runs the full NN on the frames selected by the cheap NN (it already runs the cheap NN at ingestion).
The comparison between \sys{}'s query cost and that of Focus depends on two  factors: (i) the frame selectivity $f$ and (ii) the ratio $\alpha$ between the full NN speed and the cheap NN speed.
Therefore, the ratio between \sys{}'s query delay and that of Focus is given by 
$r = 1 + \alpha/f$.
With the NNs used by Focus, $\alpha = 1/48$~\cite{focus}.

When the frame selectivity is low, e.g., the queried objects are sparse in the video, \sys{}'s query delay is significantly longer (e.g., when $f = 1\%$, $r=3$).
However, as the selectivity increases, the query delay difference between \sys{} and Focus quickly diminishes, e.g., when $f=10\%$, $r=1.2$; when $f=50\%$, $r=1.04$. 
Furthermore, as the speed gap between the two NNs enlarges, e.g., with an even cheaper NN, the query delay difference quickly diminishes as well.


%% file: related.tex
\section{Related Work}
\label{sec:related}

\Paragraph{Optimizing video analytics}
Catering to retrospective video analytics, BlazeIt~\cite{blazeit} proposes a query model and corresponding execution techniques~\cite{blazeit}.
NoScope~\cite{noscope} reduces query cost with cheap early filters before expensive NN.
To run NNs on mobile devices, MCDNN~\cite{mcdnn} trades off between accuracy and resource constraints by model compression.
\Paragraph{Optimizing live video analytics}
For distributed, live video analytics, VideoStorm~\cite{videostorm} and VideoEdge~\cite{videoedge} search for best knobs and query placements over clusters to meet accuracy/delay requirements. 
For live video analytics on the edge, LAVEA\cite{lavea} and Vigil~\cite{vigil} partitions analytics pipelines between the edge and the cloud. 
Jain \textit{et al.}~\cite{jain18arxiv} optimize video analytics over multiple cameras through cross-camera correlations.
Pakha \textit{et al.}~\cite{pakha18hotcloud} co-tune network protocols with video analytics objectives, e.g., accuracy. 
However, all the systems are incapable of optimizing ingestion, storage, retrieval, and consumption in conjunction. 

\Paragraph{Video/image storage}
Facebook's Haystack~\cite{haystack} accelerates photo access through metadata lookups in main memory.
Intel's VDMS~\cite{remis17hotstorage, vdms} accelerates image data access through a combination of graph-based metadata and array-based images backed by  TileDB~\cite{tiledb}. 
They focus on images rather than videos.
Targeting NN training, 
NVIDIA's Video Loader~\cite{nvvl} (a wrapper over NVDEC and FFmpeg)
optimizes random loads of encoded video frames. 
To support video analytics at scale, Scanner~\cite{scanner} organizes video collections and raster data as tables in a data store and executes costly pixel-level computations in parallel.
All these systems are short on controlling visual data formats according to analytics. 
NVIDIA DeepStream SDK~\cite{deepstream} 
supports video frames flow from GPU's built-in decoders to stream processors without leaving the GPU. 
It reduces memory move, but no fundamental change in trade-offs between retrieval and consumption. 

\Paragraph{Time-series database}
Recent time-series data stores co-design storage format with queries ~\cite{summarystore,btrdb}. 
However, the data format/schema (timestamped sensor readings), the operators (e.g., aggregation), and the analytics structure (no cascade) are different from video analytics.
While some databases~\cite{aerospike,influxdb} provide benefits on data aging or frequent queries, they could not make storage decisions based on video queries as they are oblivious to the analytics.

\Paragraph{Multi-query optimization}
Relational databases~\cite{sellis88tods} and streaming databases ~\cite{babcock02pods, motwani03cidr} enable sharing data and computation across queries with techniques such as scan sharing~\cite{scansharing, cooperativescans, lang07cde}. 
By doing so, they reduce data move in memory hierarchy and 
coalesce computation across queries. 
\sys{}, in a similar fashion, support data sharing among multiple possible queries, albeit at configuration time instead of at run time.
By doing so, \sys{} coalesces data demands across queries/operators and hence reduces the ingestion and storage cost. 
Through load shedding~\cite{loadshedding, aurora-1, aurora-2}, streaming databases trade accuracy for lower resource consumption;
\sys{} makes similar trade-offs for vision operators.

\Paragraph{Video systems for human consumers}
Many multimedia server systems in 90's stored videos on disk arrays in multiple resolutions or in complementary layers, in order serve human clients~\cite{keeton95ms,chiueh93multimedia}. 
Since then, Kang \textit{et al.}~\cite{kang09ms} optimizes placement of on-disk video layers in order to reduce disk seek. 
Oh \textit{et al.} \cite{oh00sigmod} segments videos into shots, which are easier for humans to browse and search.
Recently, SVE~\cite{sve} is a distributed service for fast transcoding of uploaded videos in datacenters. 
ExCamera~\cite{excamera} uses Amazon lambda function for parallel video transcoding. 
These systems were not designed for, and therefore are oblivious to, algorithmic consumers. 
They cannot automatically control video formats for video analytics.

%% file: conclusion.tex
\section{Conclusions}

\sys{} automatically configures video format knobs for retrospective video analytics. 
It addresses the challenges by the huge combinatorial space of knobs, the complex knobs impacts, 
and high profiling cost. 
\sys{} explores a key idea called backward derivation of configuration: 
the video store passes the video quantity and quality desired by analytics backward to retrieval, to storage, and to ingestion. 
\sys{} automatically derives complex configurations. 
It runs queries as fast as up to 362$\times$ of video realtime.


%% file: ack.tex

\begin{acks}
The authors were supported in part by NSF Award 1619075 (including REU) and a Google Faculty Award. 
The authors thank the anonymous reviewers for their feedback and Dr. Johannes Gehrke for shepherding. 
The authors thank NVIDIA for their GPU donation. 
\end{acks}